\documentclass[aps,floatfix,prd,showpacs]{revtex4}
\usepackage{mathtools}
\usepackage{amsmath,amssymb,epsfig,bm}
\usepackage[english]{babel}
\addto{\captionsenglish}{}
\usepackage{amsfonts}
\usepackage{xcolor}
\usepackage{caption}
\usepackage{titlesec}
\usepackage{listings}
\usepackage{placeins}
\usepackage{graphics}
\usepackage{graphicx}
\usepackage{wrapfig}
\usepackage{etoolbox}
\usepackage{xcolor}

\usepackage{placeins}
\usepackage{mathrsfs}
\usepackage{setspace}
\usepackage[utf8]{inputenc}
\usepackage[titletoc,toc,title]{appendix}
\usepackage{wrapfig}
\usepackage{ dsfont }
\usepackage{subfig}
\usepackage{graphicx}
\usepackage{booktabs}
\numberwithin{equation}{section}
\titleformat{\paragraph}
{\normalfont\normalsize\bfseries}{\theparagraph}{1em}{}
\titlespacing*{\paragraph}
{0pt}{3.25ex plus 1ex minus .2ex}{1.5ex plus .2ex}

\begin{document}

\title{The cusp properties of High Harmonic Loops}

\author{Despoina Pazouli}%
\email{Despoina.Pazouli@nottingham.ac.uk}
\affiliation{%
 School of Physics and Astronomy, University of Nottingham
 \\Nottingham NG7 2RD, United Kingdom
}%
\author{Anastasios Avgoustidis}%
\email{Anastasios.Avgoustidis@nottingham.ac.uk}
\affiliation{%
 School of Physics and Astronomy, University of Nottingham
 \\Nottingham NG7 2RD, United Kingdom
}%
\author{Edmund J. Copeland}
\email{Edmund.Copeland@nottingham.ac.uk}
\affiliation{%
 School of Physics and Astronomy, University of Nottingham
 \\Nottingham NG7 2RD, United Kingdom
}%

\date{\today}

\begin{abstract}
In determining the gravitational signal of cusps from a network of cosmic strings loops, a number of key parameters have to be assumed. These include the typical number of cusps per period of string oscillation and the typical values of the sharpness parameters of left and right moving waves on the string, evaluated at the cusp event. Both of these are important, as the power stored in the gravitational waves emitted from the loops of string is proportional to the number of cusps per period, and inversely proportional to the product of the sharpness parameters associated with the left and right moving modes on the string. In suitable units both of these quantities are usually thought to be of order unity. In order to try and place these parameters on a more robust footing, we analyse in detail a large number of randomly chosen loops of string that can have high harmonics associated with them, such as one might expect to form by chopping off an infinite string in the early universe. This allows us to analyse tens of thousands of loops and obtain detailed statistics on these crucial parameters. While we find in general the sharpness parameters are indeed close to unity, as assumed in previous work (with occasional exceptions where they can become $O(10^{-2})$), the cusp number per period scales directly with the number of harmonics on the loop and can be significantly larger than unity. This opens up the possibility of larger signals than would have otherwise been expected, potentially leading to tighter bounds on the dimensionless cosmic string tension $G\mu$. 
\end{abstract}

\maketitle

\pagenumbering{arabic}
\section{Introduction}\label{Intro}

Cosmic strings are line-like topological defects produced by symmetry breaking phase transitions in a wide range of early universe models~\cite{Vilenkin:2000jqa,Kibble:1980mv,Hindmarsh:1994re,Sakellariadou:2006qs,Copeland:2009ga,Copeland:2011dx}. They form tangles or networks that evolve dynamically and can produce a host of potentially observable signals. In particular, they are active, incoherent sources of cosmological perturbations and so their predicted effects on the CMB are very different to those of passive, coherent perturbations generated by cosmic inflation~\cite{Hu:1997hp,Pen:1997ae,Albrecht:1997nt,Albrecht:1997mz,Avelino:1997hy}. This has allowed cosmic strings to be strongly constrained, having a maximum allowed contribution to the CMB anisotropy at the level of $\sim1\%$~\cite{Ade:2015xua,Charnock:2016nzm}.

While cosmic strings have been ruled out as the main source of the observed CMB anisotropy, they remain an important subject in modern cosmology. Indeed, the formation of string networks is a generic prediction in a wide range of models of the early universe~\cite{Copeland:1994vg,Dvali:1998pa,Burgess:2001fx,Jeannerot:2003qv,Kachru:2003sx,Jones:2003da,Burgess:2004kv,Baumann:2007ah}, and so they are extremely interesting from a theoretical point of view.  At the same time, they have a rich phenomenology with observational signals relevant to several areas of cosmology and astrophysics \cite{Copeland:2011dx}. Thus, cosmic strings open an exciting observational window into the early universe. Their observation would be a major discovery in Physics, and as a bonus it would also provide important quantitative information about the physics of the early universe (e.g. the energy scale of the symmetry breaking phase transition that produced the string network). On the other hand, even failure to observe strings is of significant scientific value: as observational sensitivity improves and bounds on cosmic strings become tighter, we are excluding more of the parameter space of our models of the early universe. 

The discovery of gravitational waves has reinforced interest in the physics of cosmic strings. 
The evolution of string networks leads to the production of closed string loops, which decay via gravitational wave emission. At this point we should acknowledge this is not a uniformly accepted outcome for the string decay. Being modelled as Abelian-Higgs strings, there are also claims in the literature that the dominant form of decay is via the fields themselves and not gravitationally \cite{Hindmarsh:2017qff}. We will be considering the case where the primary decay channel is through gravitational radiation, giving rise to a stochastic background of gravitational waves. Most of the emission comes from specific events on the string, known as cusps, arising when the local velocity of the string momentarily hits the speed of light, thus producing a burst of beamed gravitational radiation. Kinks, points on the string where the tangent to the string is discontinuous, are also known to contribute significantly to the gravitational wave signal. These signals are now being targeted by gravitational wave detectors including LIGO~\cite{Abbott:2009rr,Aasi:2013vna,Abbott:2017mem} and LISA~\cite{Cohen:2010xd,Auclair:2019wcv}. Such targeting brings with it the need for a better quantitative understanding of gravitational radiation from string networks. Some of the most stringent observational bounds on cosmic strings come from their predicted stochastic gravitational wave background which can be constrained either indirectly through pulsar timing observations~\cite{Bouchet:1989ck,Pshirkov:2009vb,Sanidas:2012ee,Ringeval:2017eww, Blanco-Pillado:2017rnf} or directly by gravitational wave detectors~\cite{Aasi:2013vna}. However, these constraints are also the most sensitive to largely unknown parameters, like for example the typical size of string loops in the network. Indeed, while the evolution of the long string component of the network is well understood, the quantitative modelling of the loop component remains uncertain. In particular, the typical size of loops depends on the loop production function~\cite{Vilenkin:2000jqa} (the number density of loops as a function of  loop size and time), which has been the subject of active debate in recent years~\cite{Martins:2005es,Ringeval:2005kr,Vanchurin:2005pa,Dubath:2007mf,Hindmarsh:2008dw,Hindmarsh:2017qff,Blanco-Pillado:2013qja,Blanco-Pillado:2017oxo,Auclair:2019zoz,Blanco-Pillado:2019vcs}. 

At present, there are three models used for deriving constraints and forecasts on string networks based on their stochastic gravitational wave background. These models, referred to as model 1/I, 2/II and 3/III in references~\cite{Abbott:2017mem,Auclair:2019wcv} respectively, differ significantly in their assumed/derived loop distribution functions. 
As we are entering this exciting era of direct gravitational wave detection it is imperative that as the modelling be improved, the associated loop parameters become better quantified. In this paper, we focus on two key parameters, the number of string cusps per oscillation period and the sharpness of cusps, both of which are important components of the overall gravitational signal from cusps. 
While there has been a considerable amount of work on the role of cusps on networks of cosmic string loops (see for example  \cite{Kibble:1982cb,Burden:1985md,Copeland:1986kz,Garfinkle:1987yw,Damour:2000wa,Damour:2001bk,Damour:2004kw,Blanco-Pillado:2013qja,Wachter:2016hgi,Stott:2016loe,Blanco-Pillado:2017oxo,Blanco-Pillado:2019nto,Auclair:2019wcv}), the distribution of cusps on higher harmonic loops has not to date been studied in detail. An early attempt to address the issue can be found in Copi and Vachaspati \cite{Copi:2010jw} who used numerical simulations to characterise attractor non-self intersecting loop shapes, studying their length, velocity, kink, and cusp distributions. To reach that stage they began with initial loops containing M higher harmonic modes and argued that such loops have on average $M^2$ cusps. They also discovered that on average, large loops will split into 3M stable loops within two oscillation periods (independently of M), with the stable loops being described by a degenerate kinky loop, coplanar and rectangular. These final loops were found to have a 40\% chance of containing a cusp. 

In reality, we do not really know the harmonic distribution at formation of cosmic string loops, but we do know there could be loops formed off the long string network or as individual loops in the early universe that have many harmonics on them. The traditional picture of such loops is that as they evolve, the majority of them undergo a period where they self intersect. The initial loop then breaks into two daughter loops, with the accompanying formation of a pair of kinks on each daughter loop. These may well then self intersect, and a cascade process takes place whereby the initial high harmonic loop ends up into a class of much smaller non-self intersecting loops  \cite{Scherrer:1989ha, Albrecht:1988vj, York:1989kf}. The effect of the fragmentation process on the gravitational wave production of the network was studied in \cite{Scherrer:1990pj,Casper:1995ub}. The cusps associated with such non-self intersecting loops play an important role through the strong beams of high frequency gravitational waves they produce, which leads to both a stringy non-Gaussian distribution in the stochastic ensemble of gravitational waves generated by a cosmological network of oscillating loops, but also, and crucially, it can include occasional sharp bursts of gravitational waves from the cusp regions that stand out above the confusion of gravitational wave noise made by smaller overlapping bursts. The results of Damour and Vilenkin \cite{Damour:2001bk} suggest that if only 10\% of all string loops have cusps, the gravitational wave bursts would be detectable by Adv LIGO or LISA for string tensions $\mu$ down below $G\mu \sim 10^{-13}$, where $G$ is Newton's constant. 

In determining these constraints, there are two important parameters whose values need to be assumed. Given the spacetime position of a string $X_{\mu}(\sigma, \tau)$, where $\sigma,\, \tau$ are the worldsheet coordinates of the string, as we will shortly see the general solution for the string is given in terms of right and left moving modes travelling along it, $\vec{a}(u)$ and $\vec{b}(v)$ where $u=\sigma-\tau$ and $v=\sigma + \tau$. Now two key parameters involved in the gravitational wave calculation are the second derivative of the string position evaluated at the cusp $\partial^2_{t}X$ (a measure of ``cusp sharpness"), and the average number of cusps formed per loop period $T_\ell = \ell/2$  where $\ell$ is the invariant length of the loop. In Equation (3.21) of \cite{Damour:2001bk}, the authors argue that the generic order of magnitude estimate for $|\vec{a}''| \sim 2\pi/\ell\sim |\vec{b}''|$, where $\vec{a}'' \equiv \frac{d^2 \vec{a}}{d u^2}$ and $\vec{b}''\equiv \frac{d^2 \vec{b}}{d v^2}$ evaluated at the cusp. In other words they expect the coefficient to be of order unity ($2\pi/\ell$ is the natural unit for a string loop). In terms of the number of cusps per loop oscillation, quantified by parameter $c$ in Equation (5.14) of \cite{Damour:2001bk}, the authors consider typical values of $c\sim 1$. This parameter is meant to account for the possibility that all of the loops have of order one cusp per oscillation ($c\sim1$) or, for example, only 10\% of the loops have a cusp on them per oscillation ($c\sim 0.1$). As shown in Figure 1 of  \cite{Damour:2001bk}, $c$ can have a significant impact on the strength of the gravitational wave amplitude of bursts emitted by cosmic string cusps. Similarly, knowing the true range of values of $|\vec{a}''|$ and $|\vec{b}''|$ is also important for a proper estimate of the strength of the signal emerging from cusp bursts. This is clear from Equations (3.11, 3.12) and (3.23) of  \cite{Damour:2001bk} where the logarithmic Fourier transform of the gravitational wave burst asymptotic waveform for the cusp emission, hence the amplitude of the wave arriving on Earth, depends on terms of the form $1/(|\vec{a}''||\vec{b}''|)^{1/3}$. In particular, if it turns out that a significant fraction of the cusps had associated values $|\vec{a}''|\ll 1$ and $|\vec{b}''| \ll 1$ (in units of $2\pi/\ell$) then it could lead to a significant enhancement of the strength of the signal produced.       
   
It is apparent from the above discussion that to accurately quantify gravitational radiation from string networks one must understand: (a) how these two parameters (number of cusps per oscillation period and cusp sharpness) behave as functions of the loop harmonic number and (b) what is the expected loop distribution in terms of their harmonic content, or, at the very least, what is the harmonic content of a typical loop in the network. Here, we focus on (a), considering the behaviour of individual loops in isolation, without examining the typicality of the loop in the network (we will present our results on (b) in a forthcoming publication~\cite{inprog}). A key goal of this work is to analyse these two parameters for a wide range of high harmonic loops to establish whether there is a correlation between the number of cusps and the harmonic order of the loop, and what range/distribution of values we have for the magnitude of $|\vec{a}''|$ and $|\vec{b}''|$ evaluated at the cusps. For concreteness, we work with the odd-harmonic string~\cite{Brown:1991ni, DeLaney:1989je}, a family of cosmic string solutions which can be expressed in analytic form to arbitrarily high harmonics. In the following we will use the convention that spacetime indices are in Greek, taking the values $\mu=0,1,2,3$, while space indices are in Latin taking the values $i=1,2,3$.   
   
This paper is organised as follows. In Section~\ref{NGstring} we review the basics of the Nambu-Goto approach to string modelling, while in Section~\ref{odd-harmonic} we review the odd-harmonic string solutions, which are the focus of this work. In Section~\ref{cuspoccurence} we generate large ensembles of odd-harmonic string solutions and study their cusp number and sharpness distributions.
We conclude in Section~\ref{concl}.

\section{The Nambu-Goto string}\label{NGstring}
In this section we will give a brief introduction to the dynamics of Nambu-Goto strings in flat spacetime - more details can be found for example in the classic textbooks of Vilenkin \& Shellard \cite{Vilenkin:2000jqa} and Zwiebach \cite{Zwiebach:2004tj}. 

The background spacetime where the string moves is assumed to be a smooth 4-dimensional Lorentzian manifold with a metric $g_{\mu\nu}$ and each point on the manifold is identified by the spacetime coordinates $x^{\mu}=(x^0,x^1,x^2,x^3)$. We model the closed string dynamics using the Nambu-Goto action. In this approach the string is a one-dimensional object and its world history can be represented by a two-dimensional surface in spacetime, the worldsheet. The mapping functions
\begin{equation}
X^{\mu}=X^{\mu}(\tau,\sigma)
\label{}
\end{equation}
map the worldsheet parameters $(\tau,\sigma)$, used to parametrize the two-dimensional surface, to the spacetime coordinates $x^{\mu}$. The parameter $\sigma$ describes different points on the string on and it is subject to periodic identification, since we assume a closed string, while $\tau$ is the time parameter on the worldsheet. The induced metric, i.e. the metric induced on the worldsheet from the spacetime metric $g_{\mu\nu}$, is
\begin{equation}
\gamma_{AB}=g_{\mu\nu}\frac{\partial X^{\mu}}{\partial\xi^A}\frac{\partial X^{\nu}}{\partial \xi^B}.
\label{}
\end{equation}
The indices $A,\,B$ take values from 0 to 1, and $\xi^0=\tau,\,\xi^1=\sigma$.
The Nambu-Goto action is proportional to the surface area swept out by the string in spacetime
\begin{equation}
S=-\mu \int d\tau\oint d\sigma \sqrt{-\text{det}\gamma},
\label{action}
\end{equation}
where $\mu$ is the string tension and $\text{det}\gamma$ is the determinant of the induced metric $\gamma_{AB}$.

Restricting our analysis to a cosmic string living in a flat spacetime, the spacetime metric becomes $g_{\mu\nu}=\eta_{\mu\nu}=\text{diag}\left(-1,1,1,1\right)$. By varying the action \eqref{action} with respect to the mapping functions $X^{\mu}$, we obtain the equations of motion for the string in flat spacetime
\begin{equation}
\partial_{A}\left(\sqrt{-\text{det}\,\gamma}\gamma^{AB}\partial_{B}x^{\mu}\right)=0,
\label{eom}
\end{equation}
where $\partial_A=\partial/\partial\xi^A$.
Since the action is invariant under arbitrary reparametrizations of the worldsheet, we are free to impose two conditions on the worldsheet parameters, thereby fixing the gauge. It is convenient to choose the conformal gauge $\gamma_{AB}=\sqrt{-\text{det}\gamma}\, \eta_{AB}$, where $\eta_{AB}=\text{diag}(-1,1)$, the 2 dimensional Minkowski metric. The conformal gauge imposes the Virasoro constraints
\begin{equation}
\eta_{\mu\nu}\frac{\partial X^\mu}{\partial \tau}\frac{\partial X^{\nu}}{\partial\sigma}=0,
\label{}
\end{equation}
\begin{equation}
\eta_{\mu\nu}\frac{\partial X^\mu}{\partial \tau} \frac{\partial X^\nu}{\partial \tau}+\eta_{\mu\nu}\frac{\partial X^{\mu}}{\partial\sigma} \frac{\partial X^{\nu}}{\partial\sigma}=0.
\label{}
\end{equation}
Equation \eqref{eom} is then simplified to a two dimensional wave equation of motion for the string in flat spacetime
\begin{equation}
\left(\frac{\partial^2}{\partial{\sigma}^2}-\frac{\partial^2}{\partial{\tau}^2}\right)X^{\mu}(\tau,\sigma)=0,
\label{eom2}
\end{equation}
constrained by the Virasoro conditions. The general solution of the Nambu-Goto equations of motion in the conformal gauge is therefore a superposition of a left-moving and a right-moving wave forms. Notice that a solution to \eqref{eom2} for $\mu=0$ is $X^0=\tau$, which we can choose to fix the remaining gauge invariance. Note that our gauge choices have fixed the possible reparametrizations of the worldsheet parameters, apart from a constant shift in $\sigma$. We can now write the general solution of the string motion as
\begin{equation}
X^0=\tau,\, X^i(\tau,\sigma)=\frac{1}{2}\left(a^i(u)+b^i(v)\right),
\label{motion}
\end{equation}
constrained by 
\begin{equation}
\eta_{ij}\frac{da^i}{du}\frac{da^j}{du}=1,\,\eta_{ij}\frac{db^i}{dv}\frac{db^j}{dv}=1.
\label{VirC}
\end{equation}
To reiterate, in the above we have defined $u=\sigma-\tau$ and $v=\sigma+\tau$,  $\vec{a}(u)$ and $\vec{b}(v)$ are the right- and left-moving vector mode functions respectively. Using the notation $r'(u)=dr/du$ etc, and $\vec{a}^2=\eta_{ij}a^i a^j$, the Virasoro conditions are then written simply as
\begin{equation}
\vec{a}'\,^2=\vec{b}'\,^2=1 \,,
\label{VirC2}
\end{equation}
that is, the functions $\vec{a}'(u)$ and $\vec{b}'(v)$ both have constant unit magnitude. Therefore, they trace closed curves on a unit 2-sphere, which is called the Kibble-Turok sphere.
Note that up to this point we assumed that $\gamma_{AB}$ is non-degenerate. However, there are singular points on the worldsheet where $\det{\gamma}$ becomes zero. The velocity of the string momentarily reaches the speed of light at these points, which are called cusps. Another type of singular points on the worldsheet are the kinks, which occur where the derivatives of the right- and left-movers have discontinuities. These points,  namely the cusps and kinks play an important role in the evolution of the string since the gravitational radiation they emit dominates the gravitational wave signal from the string.  

We will choose for convenience the period in $\sigma$ to be $2\pi$. Since the loop is closed, the string trajectory should satisfy that $X^{\mu}(\tau,\sigma+2\pi)=X^{\mu}(\tau,\sigma)$. It can then be shown that $a^{\mu}(u)-a^{\mu}(u+2\pi)=-b^{\mu}(v)+b^{\mu}(v+2\pi)$, and also that the string trajectory $X^{\mu}(\tau,\sigma)$ has effective period $\pi$, since $X^{\mu}(\tau+\pi,\sigma+\pi)=X^{\mu}(\tau,\sigma)$. This is the cosmic string fundamental oscillation period $T_1=\pi$. For specific families of cosmic string solutions it is possible for the period to be less than $T_1$, for example when the string trajectory is invariant under a translation of a fraction of $\pi$, such as the case with a non-planar single harmonic loop \cite{Anderson:2003gg}. In the centre-of-mass-frame, the functions $\vec{a}' (u)$ and $\vec{b}'(v)$ should satisfy the conditions
\begin{equation}
\int_0^{2\pi} \vec{a}' du=\int_0^{2\pi} \vec{b}' dv=0.
\label{com}
\end{equation}
Therefore, in the centre-of-mass frame, the functions $\vec{a} (u)$ and $\vec{b} (v)$ have period $2\pi$, and so do $\vec{a}' (u)$ and $\vec{b}'(v)$. 

\section{The odd-harmonic string}\label{odd-harmonic}
The general solution of the cosmic string loop equations of motion can be expanded as a Fourier series. In our study we will require solutions of equations that involve the derivatives of the left- and right-movers, $\vec{a}' (u)$ and $\vec{b}'(v)$. For that reason, we will parametrize these quantities hereafter, instead of the loop trajectory. The Fourier series expansions of $a'(u)$ and $b'(v)$ are written as 
\begin{equation}
\vec{a}_N'(u)=\vec{V}+\sum_{n=1}^{N}\vec{A}_n \cos(nu)+\sum_{n=1}^{N}\vec{B}_n \sin(nu),
\label{aexpan}
\end{equation}
\begin{equation}
\vec{b}_M'(v)=-\vec{V}+\sum_{n=1}^{M}\vec{C}_n \cos(nv)+\sum_{n=1}^{M}\vec{D}_n \sin(nv),
\label{bexpan}
\end{equation}
where $N,\,M\in\mathds{N}$ correspond to the harmonic order of the string movers \cite{Vilenkin:2000jqa}. The constant terms of $\vec{a}_N'(u)$ and of $\vec{b}_M'(v)$ are constrained to be opposite to each other due to the periodicity of the loops. The solutions \eqref{aexpan} and \eqref{bexpan} must satisfy the Virasoro conditions \eqref{VirC}, which impose a non-linear set of conditions on the vector coefficients, $\vec{A}_n$, $\vec{B}_n$, $\vec{C}_n$, $\vec{D}_n$ and $\vec{V}$. A method to solve these conditions was introduced by Brown et al., who used a product representation method (and its corresponding spinorial representation) allowing them to describe the loop trajectory in terms of a matrix product \cite{Brown:1991ni, DeLaney:1989je}. Introducing $\vec{h}'_N(u)$ to represent either the derivatives of the left ($\vec{a}'(u)$) or right ($\vec{b}'(v)$) moving modes of a string with $N$ harmonic modes, they can be written in the functional form 
\begin{equation}
\vec{h}'_N(u)=\rho_{N+1}R_{z}\left(u\right)\ldots\rho_3 R_z\left(u\right)\rho_2 R_{z}\left(u\right)\rho_1 \vec{k}
\label{gNhs}
\end{equation}
where 
\begin{equation}
\rho_i=\rho\left(\theta_i,\,\phi_i\right)=R_z\left(-\theta_i\right)R_x\left(\phi_i\right)R_z\left(\theta_i\right),
\label{}
\end{equation}
such that 
\begin{equation}
0\leq \phi_i\leq 2\pi, \, 0\leq \theta_i\leq \pi ~~~~~{i=1,\, ...,\, N+1}  
\label{}
\end{equation}
and
\begin{equation}
R_z(\omega)=\begin{pmatrix}
    \cos \omega & -\sin \omega & 0 \\
    \sin \omega & \cos \omega & 0 \\
    0 & 0 & 1    
\end{pmatrix},
\label{Rz}
\end{equation}

\begin{equation}
R_x(\omega)=\begin{pmatrix}
    1 & 0 & 0 \\
    0 & \cos \omega & -\sin \omega \\
    0 & \sin \omega & \cos \omega   
\end{pmatrix}.
\label{Rx}
\end{equation}
Note that we have assumed a 3 dimensional Cartesian coordinate system xyz, with unit vectors $(\vec{i},\vec{j},\vec{k})$. We have introduced a matrix notation, where $R_x(\phi)$ denotes a rotation  of angle $\phi$ around the x axis. In expression \eqref{gNhs}, we chose the $\vec{k}$ unit vector, without loss of generality; any other unit vector can be chosen and the same rule applied, or equivalently we can change the coordinate system alignment.

The product representation method provides a general expression for the cosmic string solution up to any finite order of harmonics as a product of matrices. However, if we wish to describe the string in its centre-of-mass frame, then the cosmic string solution must also satisfy the condition \eqref{com}, which implies that both $\vec{a}'(u)$ and $\vec{b}'(v)$ have period $2\pi$. 
Although a general solution for any order $N$ that solves \eqref{com} has not been found analytically, strings comprised of only odd harmonics are in fact a solution as originally shown by Siemens and Kibble \cite{Siemens:1994ir}. This loss of generality is the small price we have to pay in order to move to the centre-of-mass frame of the cosmic string. By applying Eq.~(\ref{gNhs}) to the case where all harmonics on the $N$ harmonic string are odd, we obtain 
\begin{equation}\label{hN-odd}
\vec{h}'_N(u)=\rho_{N+2}R_z(2u)\rho_N R_z(2u)\ldots R_z(2u)\rho_3 R_z(u)\rho_1 \vec{k},
\end{equation}
with $\theta_1=-\pi/2$ and $\phi_1=\pi/2$ for both $\vec{a}'$ and $\vec{b}'$. As we have mentioned, a key goal of ours in this work is to determine the number of cusps produced per period on an odd-harmonic cosmic string, the relative positions of the points where the cusps occur and the magnitude of the second derivatives of the left and right movers evaluated at the cusps. Given that these quantities are invariant under orientations of the string in the plane, we can further simplify our solution, while making sure that the relative orientation freedom of $\vec{a}'(u)$ and $\vec{b}'(v)$ is preserved. Therefore, we are free to eliminate parameters from the orientation freedom of the string by removing the $\theta_{N+2}$ parameter from $a'(u)$ and the $\theta_{M+2}$ and $\phi_{M+2}$ parameters from $b'(v)$. Leaving the parameter $\phi_{N+2}$ in $a(u)$ ensures the relative freedom of the string movers \cite{Brown:1991ni,Siemens:1994ir}. The solutions for the derivatives of the right- and left-movers that we obtain are
\begin{equation}
\vec{a}'_N(u)=R_x(\phi_{N+2})R_z(2u)\rho_N R_z(2u)\dots R_z(2u)\rho_3 R_z(u)\vec{i}
\label{aodd}
\end{equation}
and
\begin{equation}
\vec{b}'_M(v)=R_z(2v)\rho_M R_z(2v)\dots R_z(2v)\rho_3 R_z(v)\vec{i},
\label{bodd}
\end{equation}
respectively. We will call the string appearing in equations \eqref{aodd} and \eqref{bodd}, the $N/M$ odd harmonic string, referring to the harmonic order we have chosen for $a'(u)$ and $b'(v)$ respectively. In our study we will use this parametrization of cosmic string loops to obtain our key results. The angles that appear in the matrix $\rho_i=\rho (\theta_i,\phi_i)$ will be denoted as ($\theta_{ia}, \phi_{ia})$ and $(\theta_{ib},\phi_{ib})$ for the $\rho_i$ matrices appearing in $\vec{a}'(u)$ and $\vec{b}'(v)$ respectively. It is important to note that the fact we can choose the angles at random means these are huge classes of independent solutions that we are free to analyse in order to obtain the statistics associated with the distribution of the cusps we are aiming for. By moving to the centre-of-mass frame, we have induced an extra symmetry to the cosmic string loop, which now satisfies $\vec{X}(\tau,\sigma+\pi)=-\vec{X}(\tau,\sigma)$. 

\begin{figure}%
    \centering
    \subfloat[The curves $\vec{a}'(u)$ (red) and $-\vec{b}'(v)$ (blue) on the sphere.]{{\includegraphics[scale=0.35]{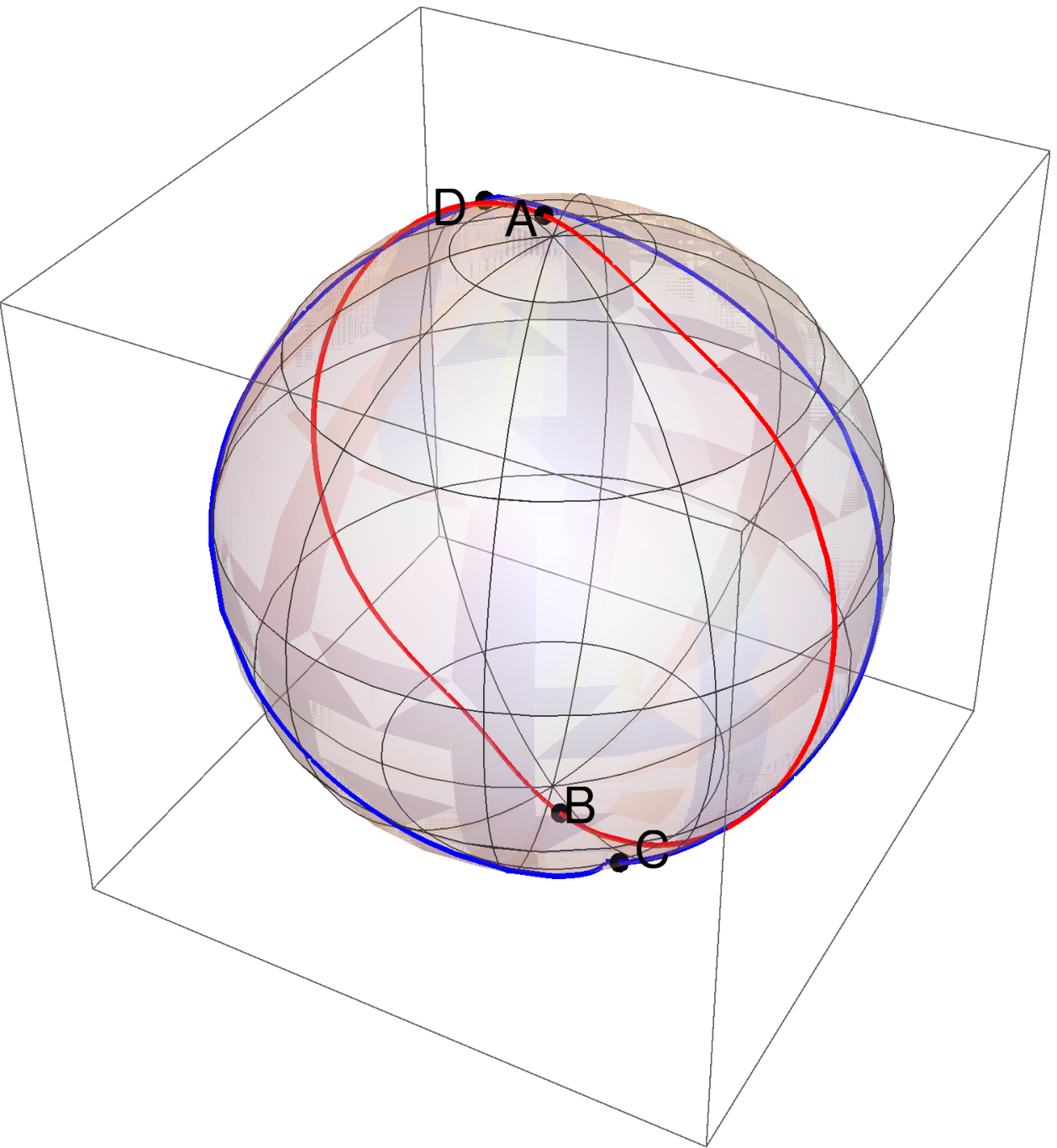} }}%
    \qquad
    \subfloat[The values of $\theta$ and $\phi$ as $\vec{a}'(\theta(u),\phi(u))$ and $-\vec{b}'(\theta(v),\phi(v))$ move on the sphere for N=3.]{{\includegraphics[scale=0.45]{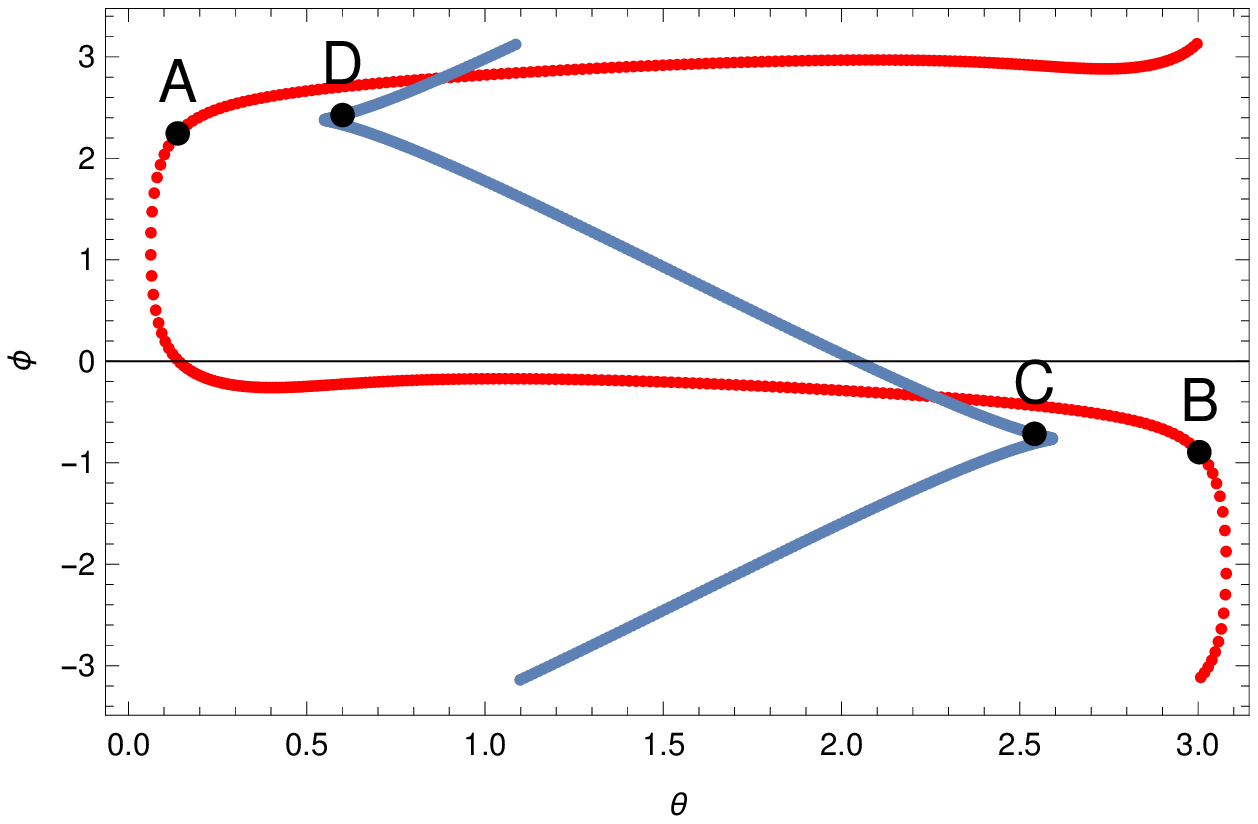} }}%
     \qquad
   \subfloat[The values of $\theta$ and $\phi$ as $\vec{a}'(\theta(u),\phi(u))$ and $-\vec{b}'(\theta(v),\phi(v))$ move on the sphere for N=13.]{{\includegraphics[scale=0.45]{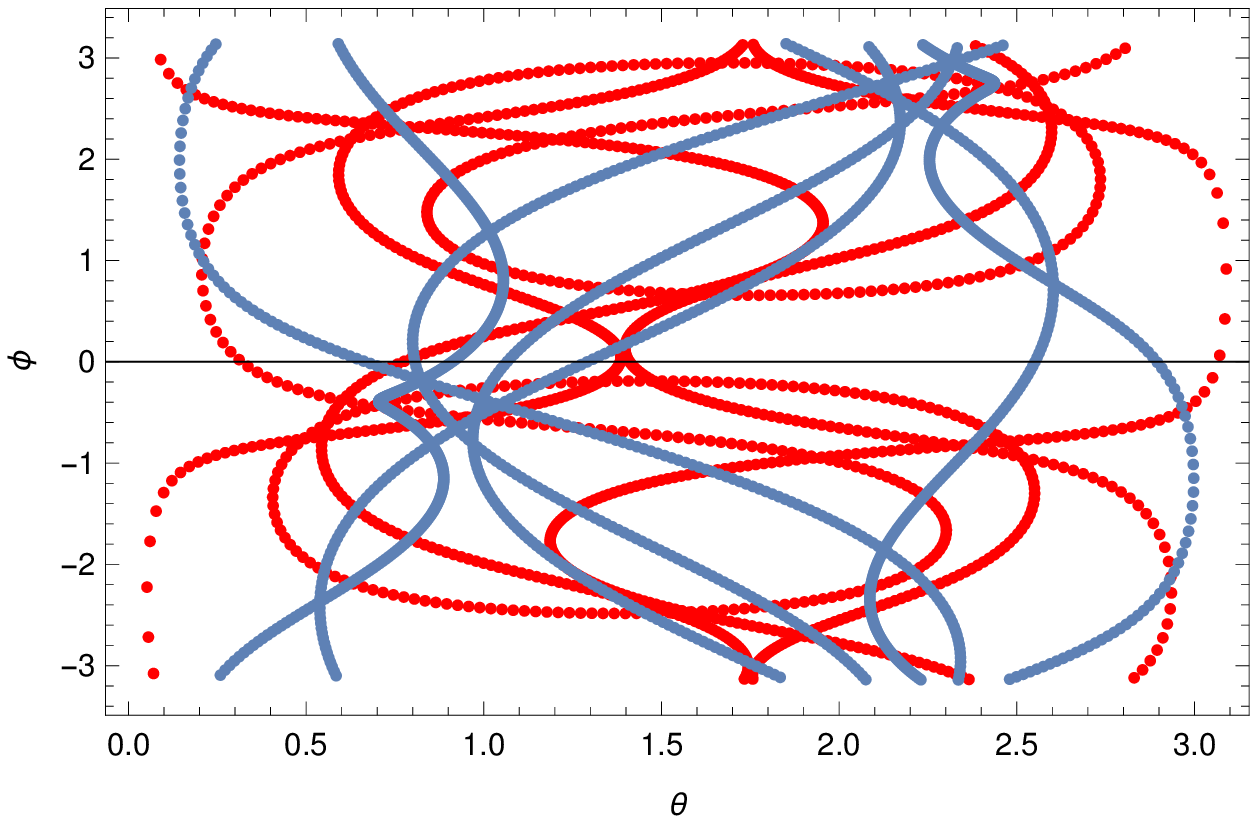} }}%
     
    \caption{Plots of the $\vec{a}'(u)$ and $\vec{b}'(v)$. In plots (a) and (b) we see the harmonic order N=3 case. The points depicted in the plots correspond to $A=(\theta_{\vec{a}'}(u=0),\phi_{\vec{a}'}(u=0)),\, B=(\theta_{\vec{a}'}(u=\pi),\phi_{\vec{a}'}(u=\pi)),\, C=(\theta_{\vec{b}'}(v=0),\phi_{\vec{b}'}(v=0)),\, D=(\theta_{\vec{b}'}(v=\pi),\phi_{\vec{b}'}(v=\pi)).$}%
    \label{}%
\end{figure}

\section{Cusp occurence}\label{cuspoccurence}
A generic property of cosmic strings is that points on the string can momentarily reach the speed of light. Differentiating equation \eqref{motion} with respect to time, we find that
\begin{equation}
\dot{\vec{X}}(\tau,\sigma)=\frac{1}{2}\left(-\vec{a}'(u)+\vec{b}'(v)\right).
\label{}
\end{equation}
A point on the string reaches the speed of light when $|\dot{\vec{X}}|^2=1$, or equivalently when
\begin{equation}
\vec{a}'(u)=-\vec{b}'(v) \Leftrightarrow \vec{a}'(u)\cdot \vec{b}'(v)=-1.
\label{cuspeq}
\end{equation}
The solutions of equation \eqref{cuspeq} are called cusps, and their location on the string worldsheet will be denoted as $(u_{\textrm{c}},v_{c})$. Schematically, these solutions occur when $\vec{a}'(u)$ and $-\vec{b}'(v)$ intersect on the Kibble-Turok sphere. Since the curves $\vec{a}'(u)$ and $\vec{b}'(v)$ are periodic, the cusp solutions $(u_{c},v_{c})$ will also occur periodically. Smooth strings, arising from the Nambu-Goto action, will generically have cusp points. However, other types of strings can exist, with discontinuities in $\vec{a}'(u)$ and $\vec{b}'(v)$, called kinks. In this type of strings, cusps are more rare since the discontinuities make intersections on the Kibble-Turok sphere less likely to occur \cite{Vilenkin:2000jqa}.

We wish to track down the occurrence of the cusps in our class of odd-harmonic strings. 
Unfortunately it is clear that our key quantities \eqref{aodd} and \eqref{bodd} are non-linear in $u,v$ and the equation \eqref{cuspeq} cannot be solved analytically in general. Hence we will solve it numerically to obtain both the number of cusps formed per period, and, by differentiating Eqns. \eqref{aodd}, \eqref{bodd} and evaluating them at $(u_{c},v_{c})$, we will also obtain all the corresponding values for $\vec{a}''(u_{c})$ and $\vec{b}''(v_{c})$ respectively, another key quantity. Fortunately, there are in fact specific choices of parameters that do lead to analytic solutions for odd-harmonic strings. This is very useful as it will allow us to test our numerical algorithm, and so we now turn to these specific cases. 
\subsection{Analytic cases}\label{AnCas}
An example of string solutions where the occurrence of cusps can be found analytically is the Kibble-Turok family of strings \cite{Kibble:1982cb}, which is a subset of the odd-harmonic string for certain choices of parameters in $\vec{a}_3'(u)$ and $\vec{b}_3'(v)$. They can be obtained from \eqref{aodd} and \eqref{bodd} if we set $\theta_{3a}=0$, $\phi_{5a}=0$ and $\theta_{3b}=0$ and $\phi_{3b}=\pi$. The only free parameter is $\phi_{3a}$, which we allow to range in the interval $[0,\pi]$. If we set $\alpha=\cos^2\left(\phi_{3a}/2\right)$, then the right- and left-movers are 
\begin{align}
\vec{a}'(u)=&\left[(1-\alpha)\cos{(u)}+\alpha \cos{(3u)}\right]\vec{i}\\
&+\left[(1-\alpha)\sin{(u)}+\alpha \sin{(3u)}\right]\vec{j}\\
&+2\sqrt{\alpha(1-\alpha)}\sin{(u)}\,\vec{k}
\label{kt1}
\end{align}
and
\begin{align}
\vec{b}'(v)=\cos{(v)}\vec{i}+\sin{(v)}\vec{j}.
\label{kt2}
\end{align}
The cusps positions can easily be obtained analytically for this family of loops by solving the cusp condition \eqref{cuspeq} for the above cosmic string solution, which provides a set of three equations with two unknowns. Note that the left-mover has no component in the z-axis, which restricts the possible values of the $u$ coordinate. The remaining two equations provide the possible $v$ values and constrain the choices of pairs $(u,v)$ that satisfy the system. We find that this family of string solutions supports two simultaneous cusps per period $\pi$ at the points $(\tau,\sigma)=(\pi/2,\pi/2)$ and $(\tau,\sigma)=(\pi/2,3\pi/2)$.

Another analytic solution that can be described by our analysis is the first order harmonic loop \cite{Anderson:2003gg} obtained from \eqref{aodd} and \eqref{bodd} for $N$=1, 
\begin{align}
\vec{a}'(u)=\cos{(u)}\vec{i}+\cos{\phi}\sin{(u)}\vec{j}+\sin{(\phi)}\sin{(u)}\vec{k}
\label{1h1}
\end{align}
and
\begin{align}
\vec{b}'(v)=\cos{(v)}\vec{i}+\sin{(v)}\vec{j}\,,
\label{1h2}
\end{align}
which also supports two cusps per period at $(\tau,\sigma)=(\pi/2,\pi/2)$ and $(\tau,\sigma)=(\pi/2,3\pi/2)$. The cusp positions are found by solving the set of equations from \eqref{cuspeq}, as we did for the Kibble-Turok string, by 
transforming from the coordinates ($\tau', \sigma'$) used in \cite{Anderson:2003gg} to our coordinate system, via $\tau'=\tau$, $\sigma'=\sigma+\pi/2$ and $\phi'=\phi+\pi$.
We were thus able to confirm the properties of both of these solutions with our numerical code, which we now go on to describe.

\subsection{Numerical method}\label{NUmmet}
We will now summarise the structure of the code we used to solve the non-linear system of equations \eqref{cuspeq}. It consists of two parts, one to construct the loops and another to solve the non-linear system. For simplicity in what follows we set the harmonic order of the left and right movers to be the same value $N$. 
\begin{itemize}
\item Part I
\begin{enumerate}
\item We choose the harmonic order $N$ and the number of loops M (note that this is not the harmonic order of the left movers $M$) we will analyse.
\item Adopting an iterative process we produce the $N$-order harmonic loop in $(N-1)/2$ steps, using the definition of the odd-harmonic loop, \eqref{aodd} and \eqref{bodd}. In each step of the iteration we choose randomly 2 angles for the $\vec{a}'_N(u)$ and 2 angles for $\vec{b}'_N(v)$, and in the last step we randomly choose one extra angle for each of the left- and right-movers, which are needed to eventually build the $N$-harmonic loop with $2N-1$ random angles in total. In this way we aim to sample the plane of angles, through a large number of random choices of angle parameters. 
\item{We append into lists the functions $\vec{a}'_N(u)$ and $\vec{b}'_N(v)$, their derivatives, $\vec{a}''_N(u)$ and $\vec{b}''_N(v)$, and the randomly chosen angles. The process of step 2 is repeated M times, to eventually obtain in lists all the required information for the M loops produced.} 
\end{enumerate}
\item Part II
\begin{enumerate}
\item We enter an iterative process where we choose the ith element of the lists we have produced in Part I, where i takes integer values from 1 to M, labelling the loop we are considering.
\item We divide the $(u,v)$ plane into equally sized grids and numerically solve equation  \eqref{cuspeq} to obtain the cusp solutions $(u_c,v_c)$, which are then appended in a list and tested to check that they are indeed solutions
i.e. that they satisfy \eqref{cuspeq}) by using the analytic expressions we saved from Part I. We then reduce the size of the grid and repeat the above process until no new $(u_c,v_c)$ pairs are found. After we obtain the list of cusp points for the chosen loop, we calculate $|a''(u_c,v_c)|,\,|b''(u_c,v_c)|$.
\item This process is repeated until we analyse all loops from Part I.
\end{enumerate}
\end{itemize}
\begin{figure}%
    \centering
   \includegraphics[scale=0.5]{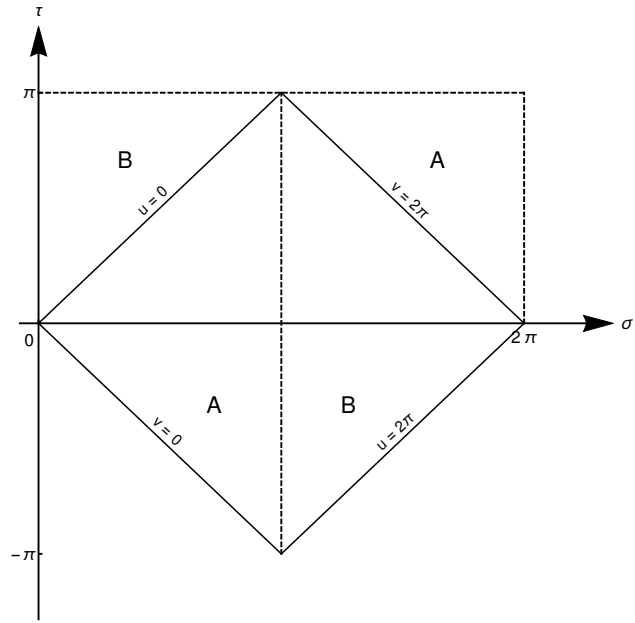} %
    \caption{The worldsheet domain in terms of the $(\tau,\sigma)$ and $(u,v)$ coordinates.}%
    \label{domains}%
\end{figure}

Since the general equation of motion for the string comprises of the periodic functions $\vec{a}(u)$ and $\vec{b}(v)$, each with a period $2\pi$, the domain in $u-v$ space $[0,2\pi)\times[0,2\pi)$ contains all the information about the string motion. In Fig.~\ref{domains}, we can see that the $[0,2\pi)\times[0,2\pi)$ $(u,v)$ domain can be mapped to the $[0,2\pi)\times [0,2\pi)$ ($\tau,\sigma$) domain, as expected. Indeed, the domains labelled A in Fig.~\ref{domains} are equivalent to each other, due to the periodicity of the string, $\vec{X}(\tau,\sigma)=\vec{X}(\tau+\pi,\sigma+\pi)$. The same holds for the domains B. 

The numerical analysis of the strings can be quite time consuming. To reduce computational time we take advantage of the extra symmetry of our string loop solutions in the centre of mass frame, $\vec{X}(\tau,\sigma+\pi)=-\vec{X}(\tau,\sigma)$, which is equivalent to $\vec{X}(u,v)=\vec{X}(u+\pi,v+\pi)$. This implies that if a cusp occurs at $(u_{c},v_{c})$, it also occurs at $(u_{c}+\pi,v_{c}+\pi)$, and we therefore only need to look for cusp solutions in the domain $[0,\pi)\times [0,2\pi)$ of the $u-v$ space, and then map them to the full domail $[0,2\pi)\times [0,2\pi)$ to obtain the full space of cusp solutions.
\subsection{Results and discussion }\label{Results-section}
In order to check on the accuracy of our simulations, we have compared our numerical results with the analytic cases mentioned in section \ref{AnCas}, namely the one-harmonic loop \cite{Anderson:2003gg} and the Kibble-Turok loop \cite{Kibble:1982cb} which can both be obtained from our odd-harmonic loops. In both cases we find the same number and positions for the cusps, with an accuracy of $10^{-5}$, and the same values for $|a_c''|,\,|b_c''|$ evaluated at the cusp. 
As a further check to our numerical method, we find that the Turok solution \cite{Turok:1984cn}, a generalization of the Kibble-Turok string with two free parameters, exhibits (generically) either two or six cusps per period, with two of them always occurring at $(\tau,\sigma)=(\pi/2,\pi,2)$ and $(\tau,\sigma)=(\pi/2,3\pi,2)$, as expected. Since the Turok string is not a subfamily of the odd-harmonic string, we input the equations of the left- and right-movers of the Turok string (given in \cite{Vilenkin:2000jqa}) directly into Part II of our code (described in section \ref{NUmmet}). Note that the Turok string can also exhibit 4 cusps per period for a specific choice of its two free parameters. Since we choose these parameters randomly, it is unlikely that we will come across this fine-tuned case. The cusp structure of the Turok string with respect to the two-parameter space is provided in \cite{Anderson:2003gg}.

\begin{figure}%
    \centering
{{\includegraphics[scale=0.55]{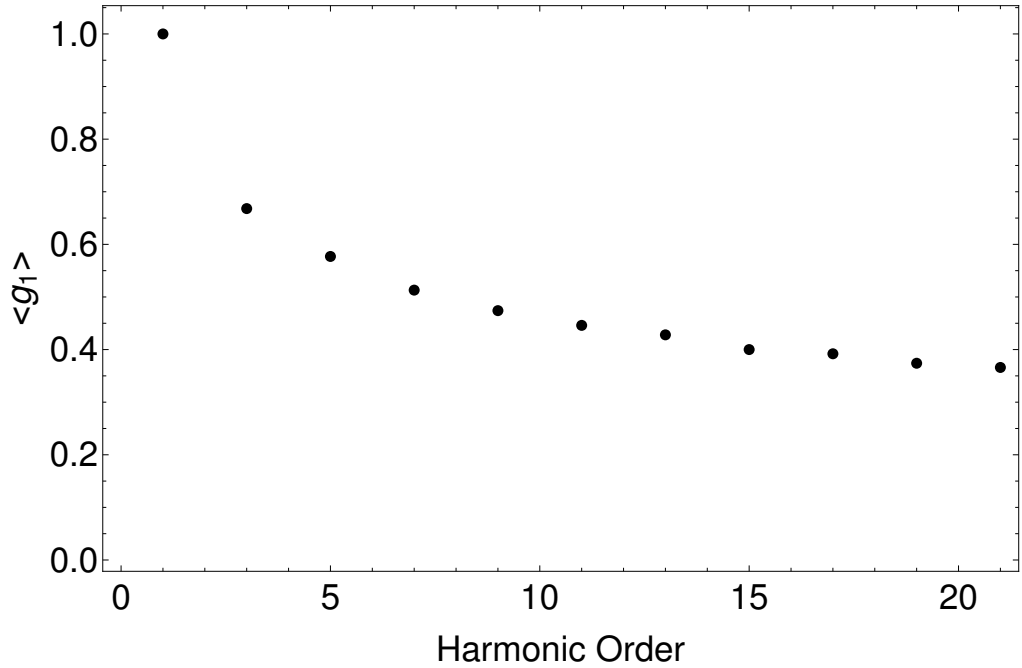} }}%
    \caption{Our numerical results for the mean value of $g_1$, with harmonic order from 1 to 21. Note that the harmonic order takes only odd values.}
\label{meang1}
\end{figure}

Having confirmed the consistency of our approach with known analytic solutions, we can confidently go on to look at cases of more general odd-harmonic strings, going up to harmonic order 21. The first question we wish to address is: what is the average value of the second derivative of the left- and right-movers? Recall this is an important contributor to the gravitational wave power emitted by cusps, and the assumption being made when calculating the associated gravitational wave power emerging from the cusp region is that in units of $2\pi/\ell$, the average value is of order 1 \cite{Garfinkle:1987yw,Damour:2001bk,Abbott:2017mem}. If it is substantially smaller, or if there are a significant number of cusps on a loop producing such small values, this will increase the associated power. Our key results are presented in Figs.~ \ref{meang1} and  \ref{histogram}.  
In Fig.~ \ref{meang1} we have calculated the mean value for the second derivative of the left- and right-movers on the cusps defined through $g_1$ which was introduced in  \cite{Damour:2001bk}
\begin{equation}\label{g1-def}
g_1=\left(|\vec{a}''(u_c,v_c)||\vec{b}''(u_c,v_c)|\right)^{-\frac{1}{3}},
\end{equation}
for large numbers of loops as a function of the harmonic order, ranging from $N=1$ to $21$. This is so important because the gravitational wave amplitude is proportional to $g_1$.  
The error bars indicate the usual standard error associated with the mean. They are vanishingly small for low harmonics and increase for larger harmonics, because we are not able to analyse as many large harmonic loops due to time constraints associated with analysing such complex loops.
Of particular note is the fact that $\langle g_1\rangle$ decreases rapidly initially as the harmonic order increases but eventually plateaus for large $N$. We also note that it is indeed consistent with the claims in \cite{Damour:2001bk}, namely that it is a number of order unity, given that it ranges from 1, for small $N$, down to 0.4, for large $N$. 

What about the range of possible values of $g_1$ for a given harmonic? In particular, how large can it go, and how frequent are these large values? We address this question in Fig.~ \ref{histogram}, where each of the four plots represents the frequency distribution of the parameter $g_1$ produced from 1000 loops of a specific harmonic order $N$. On the horizontal axis we have the values of $g_1$ and on the vertical axis the number of times the parameter $g_1$ obtains a value which lies in the corresponding bin. Note that the first plot, depicting the frequency distribution of $g_1$ for $N=3$ harmonic loops, decreases almost monotonically from its initial high value in the $[0.45,0.5)$ bin, except for a secondary subsidiary peak in the $[0.95,1)$ bin. As we increase the number of harmonics $N$ on the loop, the histograms become more peaked around small values of $g_1 \sim 0.4$, a feature that is particularly noticeable for the cases $N=11$ and $N=19$.
We also notice that the total number of counts increases as we increase the harmonic order, $N$, of the loops. This occurs because the average number of cusps per period increases as we increase the harmonic order. 
We can understand the typical smallest value $g_1$ takes. For the case of the $N=3$ harmonic order loop, we can show analytically (from Eqns.~$\eqref{aodd}$ and $\eqref{bodd}$) that it is 0.5, matching the numerical result. 
Unfortunately, analytic approaches break down for higher orders, as the function $g_1(u,v)$ quickly becomes complicated, and finding its maximum analytically becomes progressively harder.
It is clear from the four cases depicted in Fig.~\ref{histogram} that the number of loops with large values of $g_1 \gg 1$ become negligible, hence increasing the number of harmonics does not apparently have a significant impact in the range of possible values of $g_1$; they remain stubbornly close to the assumed value of order unity.

\begin{figure}    
    {{\includegraphics[scale=0.55]{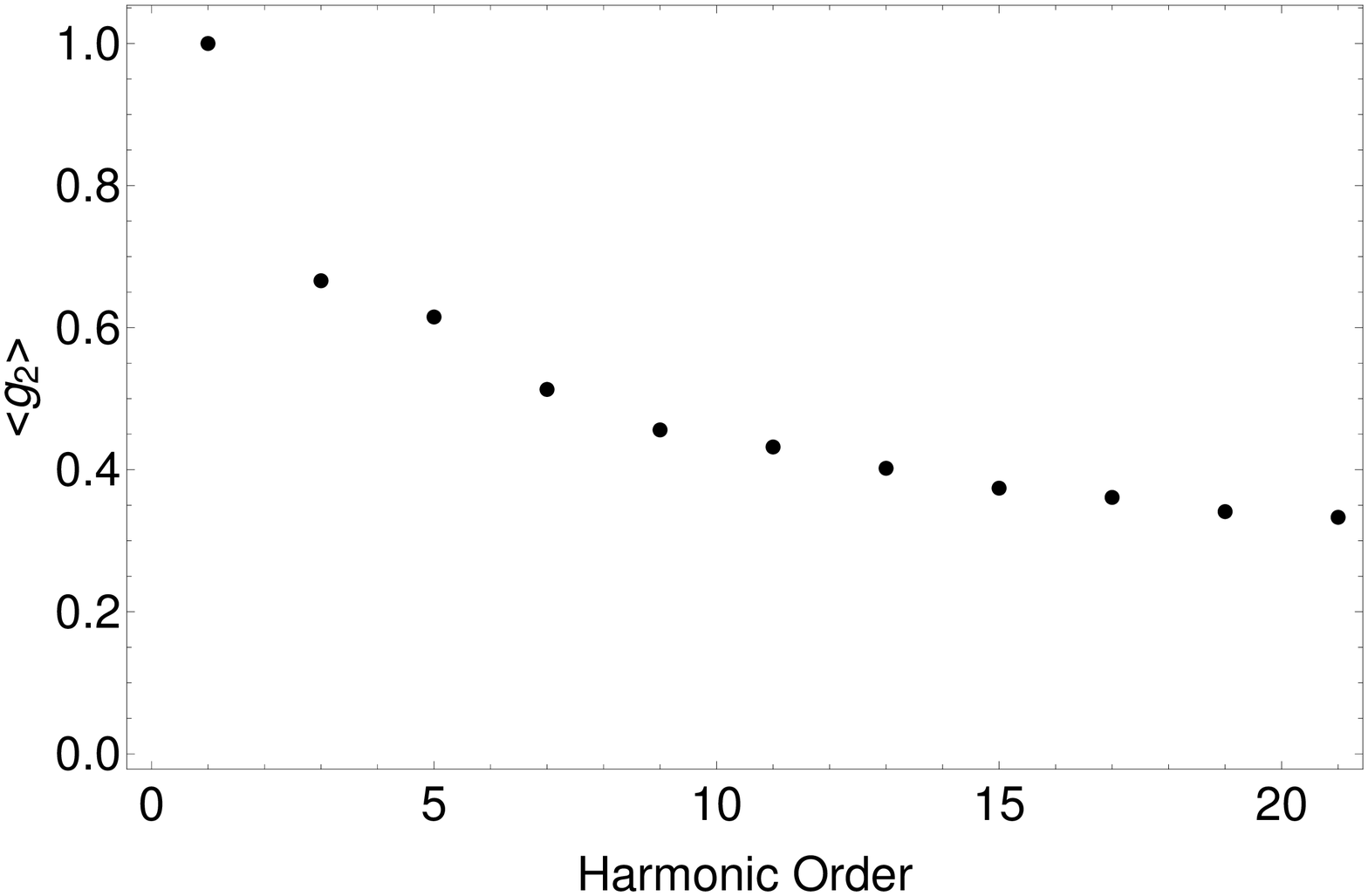} }}%
    \caption{Our numerical results for the mean value of $g_2$, with harmonic order from 1 to 21. Note that the harmonic order takes only odd values.}%
    \label{meang2}%
\end{figure}

In Fig.~\ref{meang2} we plot the mean value of yet another quantity that is related to the gravitational wave signal emitted from cusps on cosmic strings. This is given by parameter $g_2$ , defined  in  \cite{Damour:2001bk} as
\begin{equation}\label{g2-def}
g_2=\left({\rm min}\left(|\vec{a}^{\,''}|,|\vec{b}^{\, ''}|\right)\right)^{-1},
\end{equation} 
which is inversely proportional to the beaming angle of the cusp $\theta^{div}$. In particular, we define the angle $\theta^{div}$ to be the angle that divides the observation angles of a cusp into two sets; one where the gravitational wave signal is roughly the same as it is along the direction of the cusp emission ($\theta<\theta^{div}$), and one where the signal is smoothed ($\theta>\theta^{div}$), which corresponds to the gravitational wave signal away from the cusp \cite{Damour:2001bk, Damour:2004kw}. The observer receives the gravitational wave which has emanated from the cusp on the cosmic string if and only if the observation angle with respect to the direction of the cusp satisfies $\theta<\theta^{div}$. As $g_2$ decreases, the angle $\theta^{div}$ increases, which implies that the cusp signal can be received from a broader range of observation angles, and leads to an enhanced overall gravitational wave signal from cosmic strings on Earth. From Fig.~ \ref{meang2}, we notice that the average value of $g_2$ for each harmonic order follows a pattern similar to the one in Fig.~ \ref{meang1}, starting from an average value of unity at the first order harmonic string, and gradually decreasing in value until it plateaus at just below $0.4$. Once again we note that the values of $g_2$ obtained using the odd-harmonic string do not deviate significantly from the usual assumption that its value is equal to unity \cite{Damour:2001bk, Damour:2004kw}. Given both $g_1$ and $g_2$ have basically the same dependence on $|\vec{a}^{\,''}|$ and $|\vec{b}^{\,''}|$ which typically have values of order unity, it is not surprising they have the same basic shape. 

\begin{figure}    
    {{\includegraphics[scale=0.55]{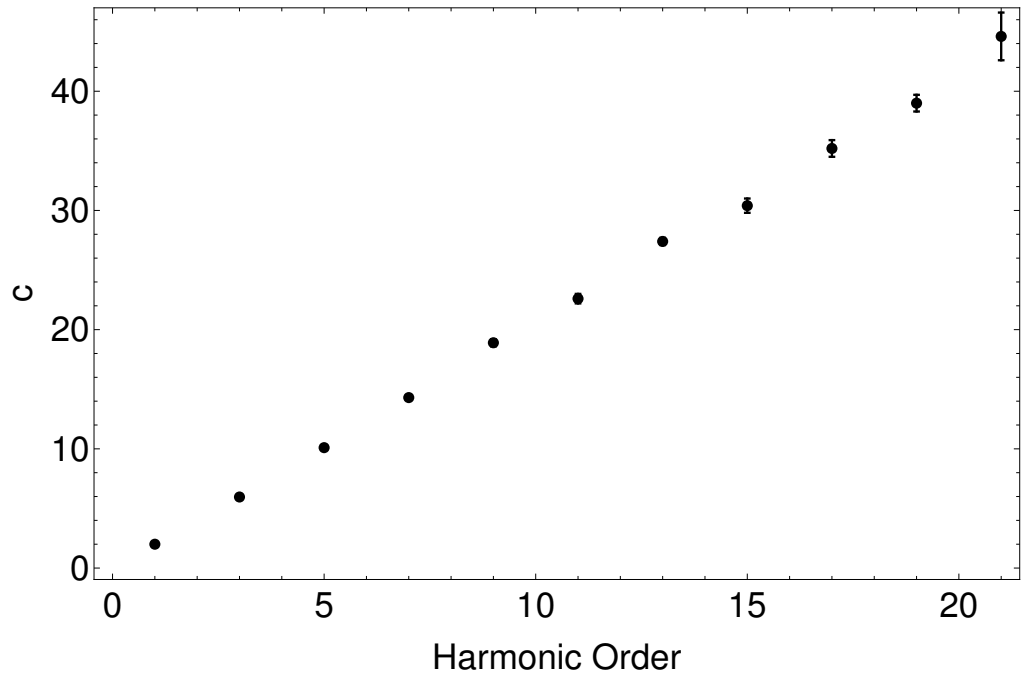} }}%
    \caption{Our numerical results for the mean value of the cusps per period $c$, with harmonic order from 1 to 21.}%
    \label{meanc}%
\end{figure}

\begin{figure}%
    \centering
    \subfloat[]
    {{\includegraphics[scale=0.6]{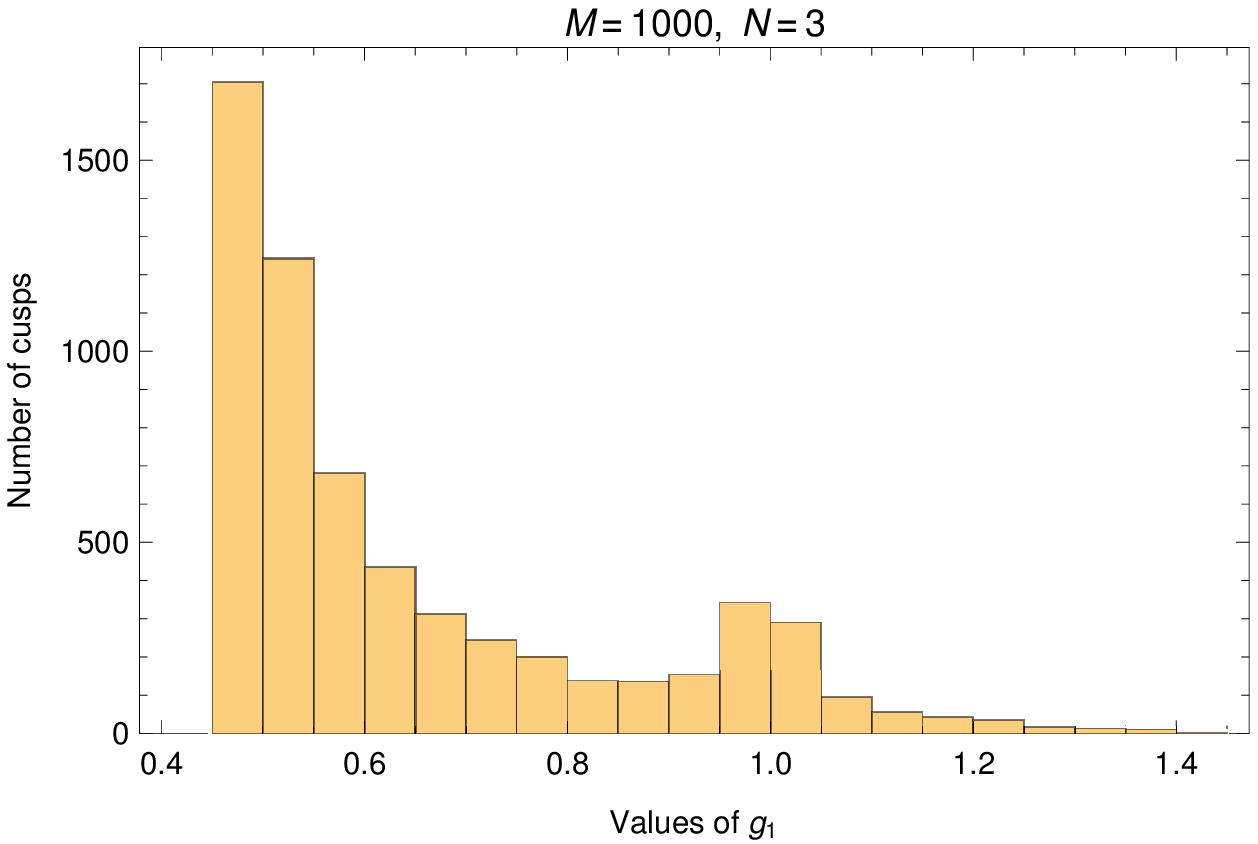} }}%
    \qquad
     \subfloat[]
    {{\includegraphics[scale=0.6]{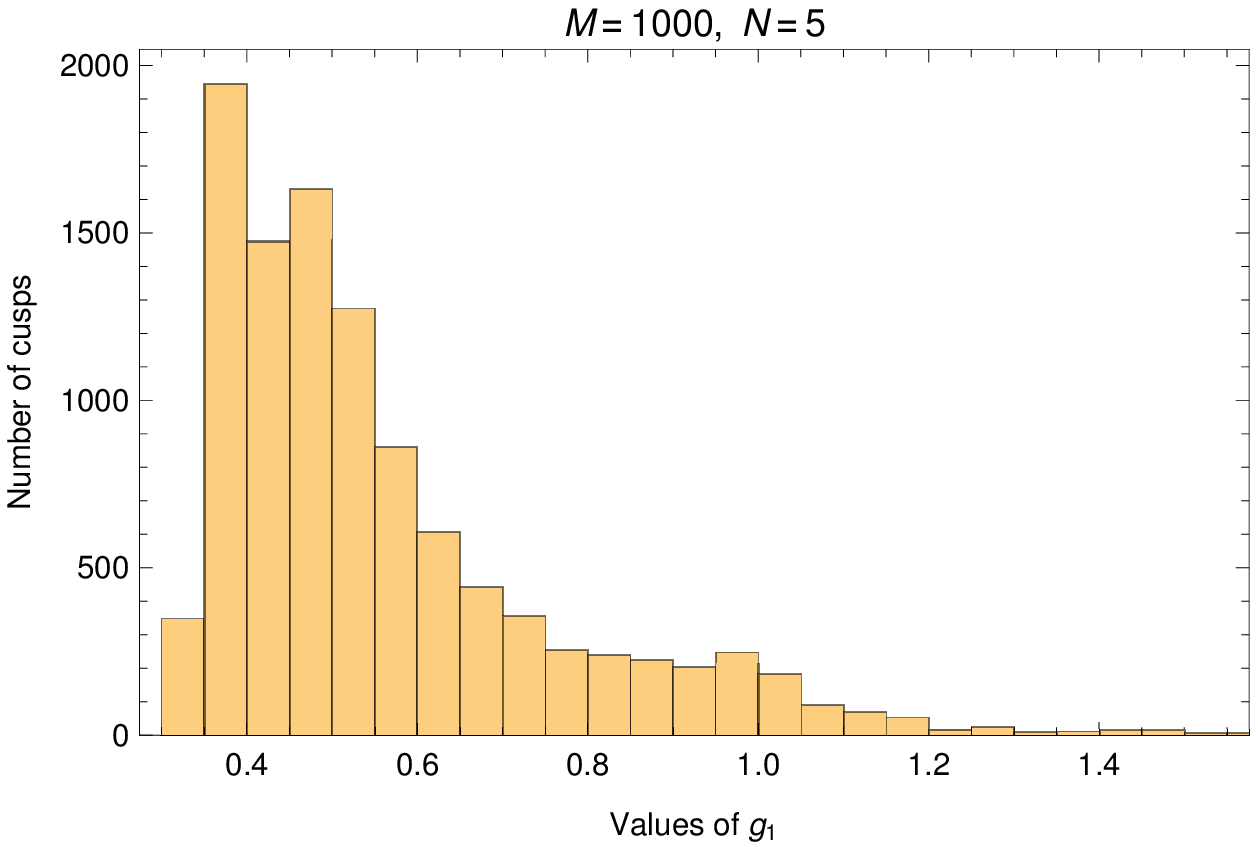} }}%
    \qquad
    \subfloat[]    
    {{\includegraphics[scale=0.6]{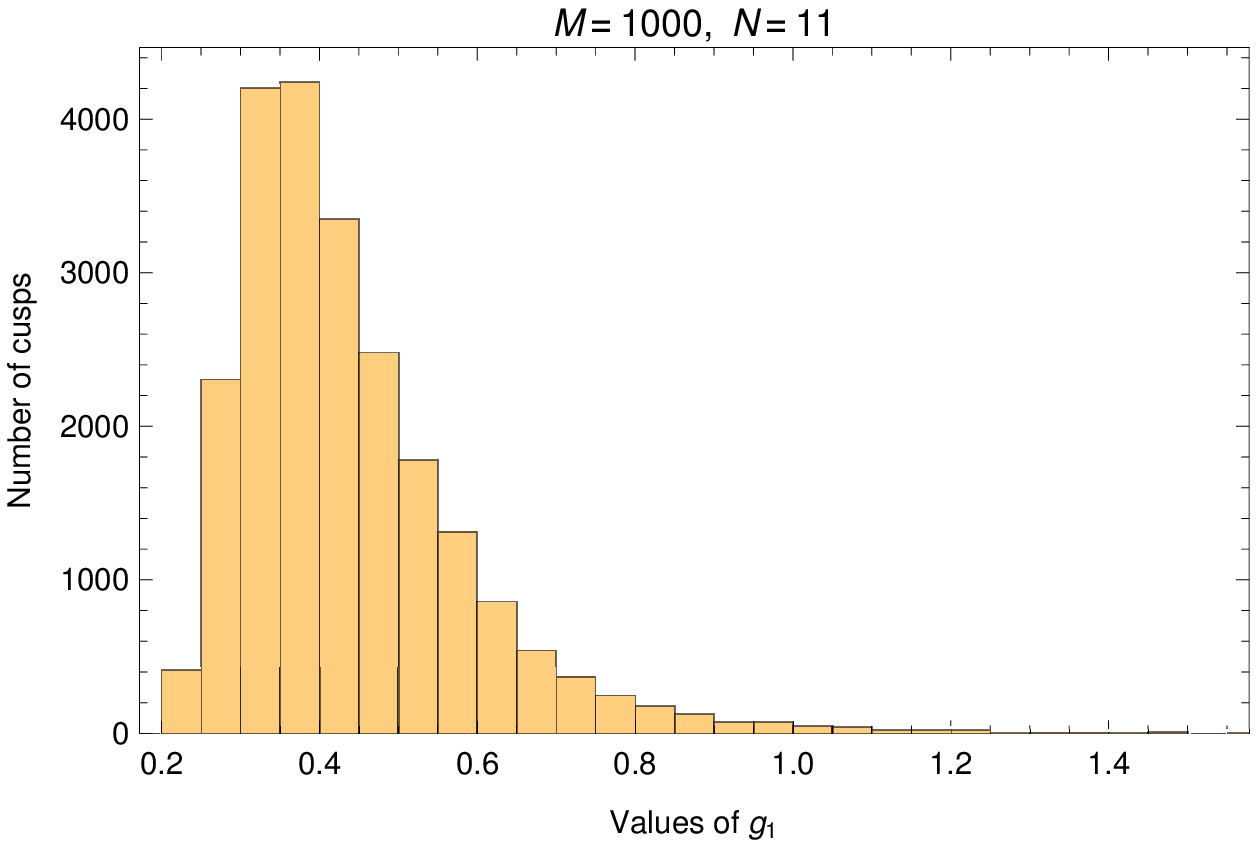} }}%
    \qquad
    \subfloat[]
    {{\includegraphics[scale=0.6]{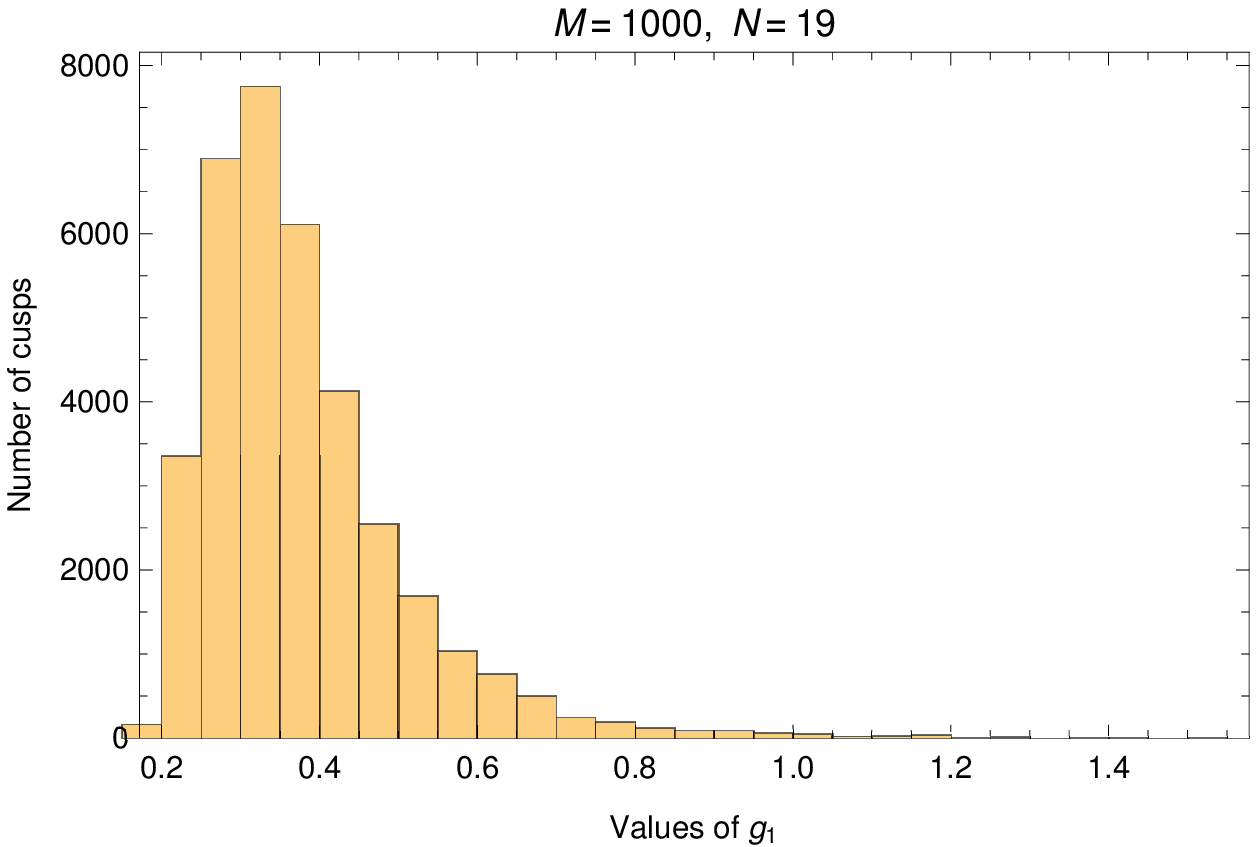} }}%
    \caption{The frequency distribution of the parameter $g_1$ calculated for different harmonic orders. Note that the data originate from M=1000 cosmic string loops for each plot. We can see that the total number of cusps increases with the harmonic order, which is expected since the average number of cusps per period for each loop also increases with the harmonic order. Note that we have cut some rare higher values of $g_1$ from the histograms. The maximum value of $g_1$ in the data is given in Table \ref{}.}%
    \label{histogram}%
\end{figure}

\begin{figure}%
    \centering
    \subfloat[]
    {{\includegraphics[scale=0.6]{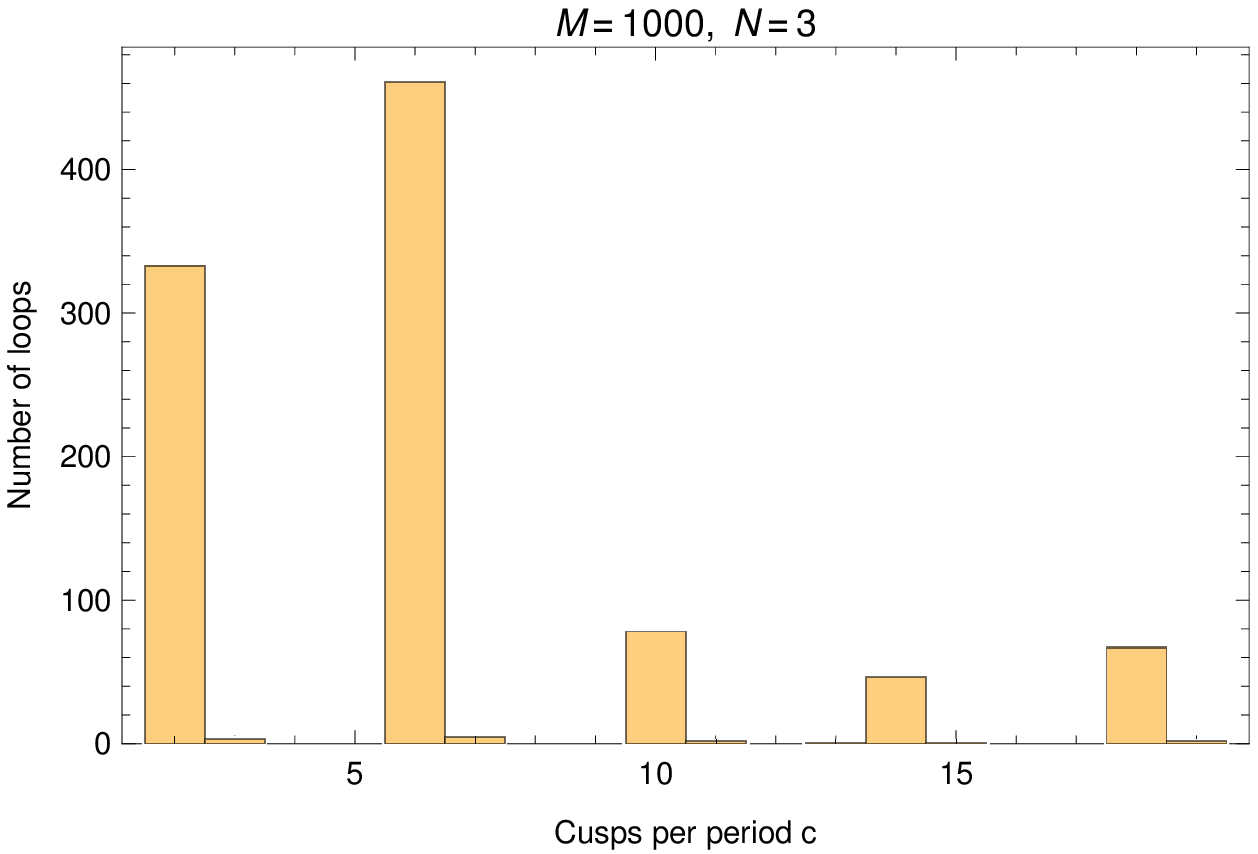} }}%
    \qquad
     \subfloat[]
    {{\includegraphics[scale=0.6]{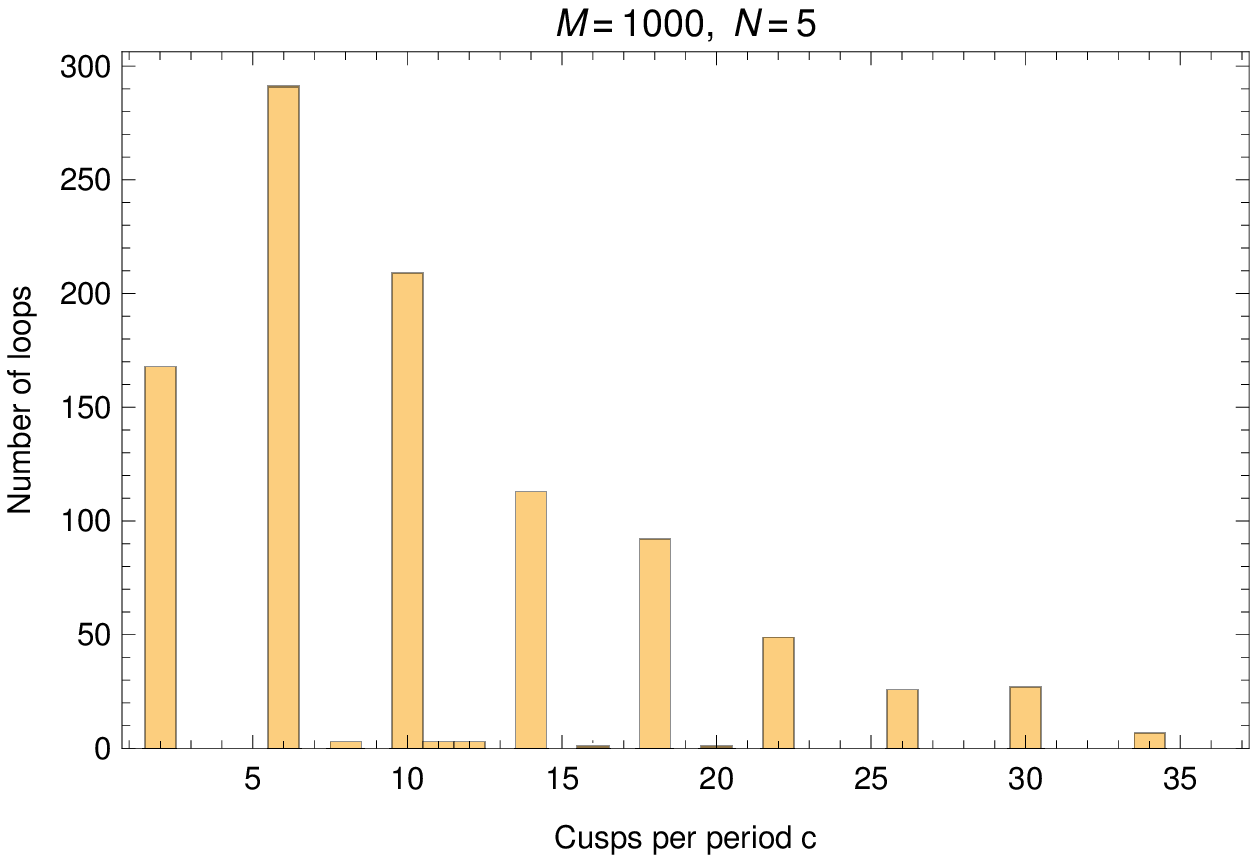} }}%
    \qquad
    \subfloat[]    
    {{\includegraphics[scale=0.6]{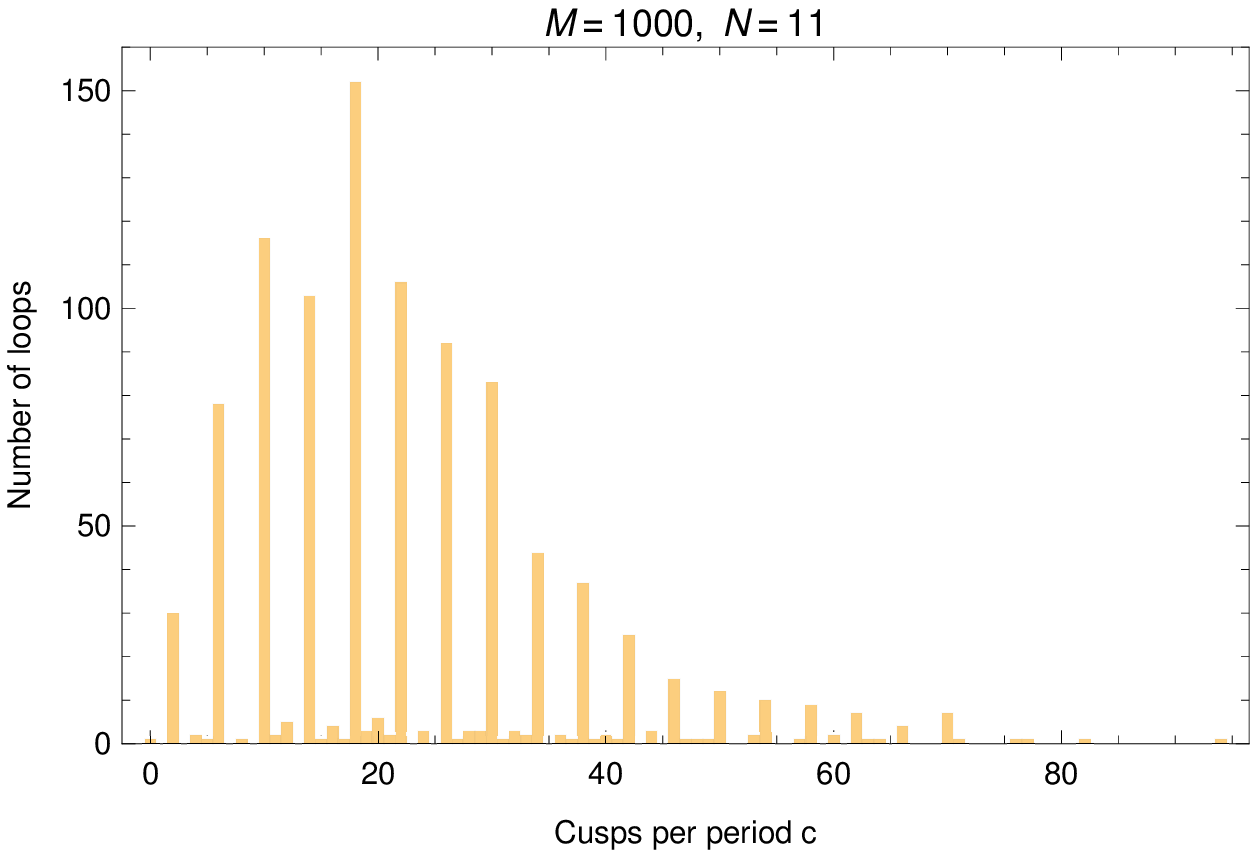} }}%
    \qquad
    \subfloat[]
    {{\includegraphics[scale=0.6]{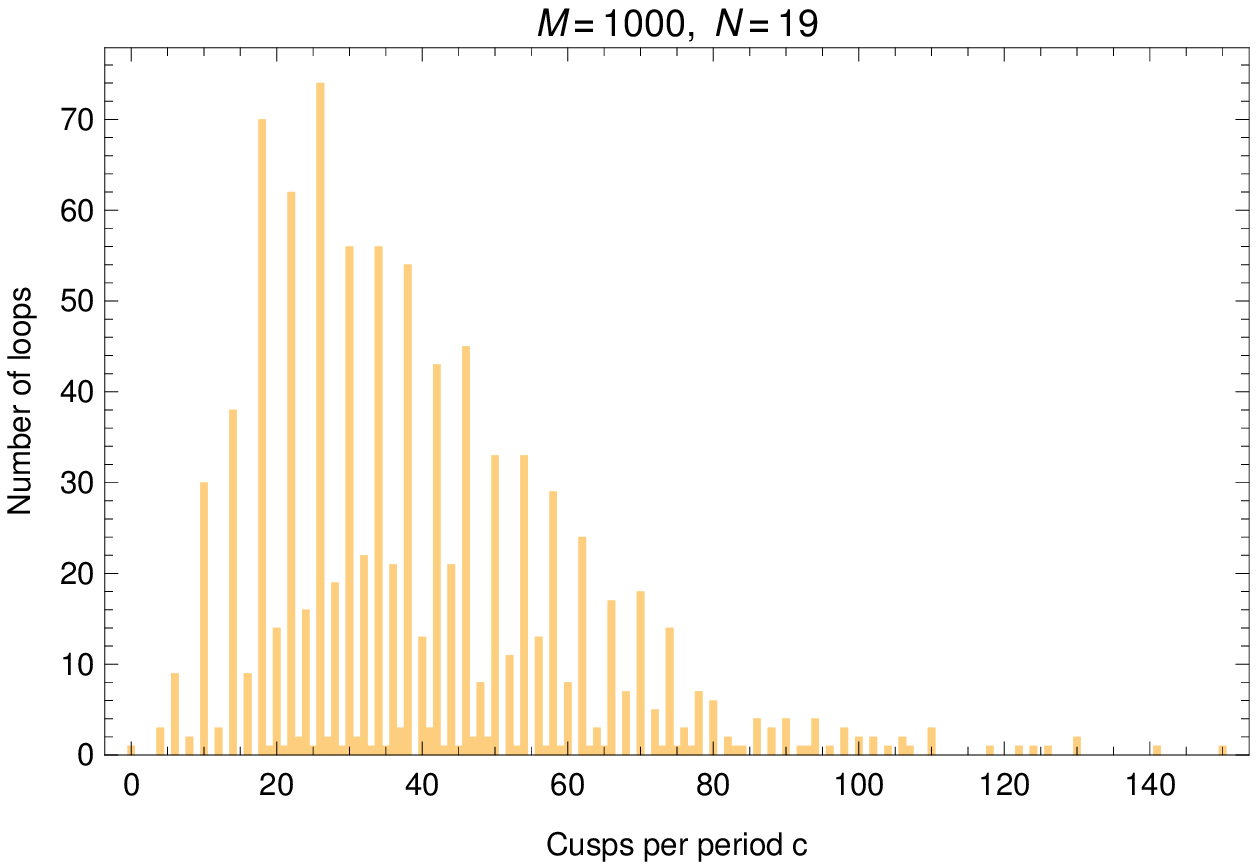} }}%
    \caption{The frequency distribution of the cusps per period. Each plot is produced from 1000 odd-harmonic string loops of the same harmonic order, N=3, N=5, N=11, N=19. The bins are $[0.5,1.5),\, [1.5,2.5),\,[2.5,3.5)$...}
    \label{histogramc}%
\end{figure}

Turning our attention to the number of cusps appearing per period on a harmonic order $N$ string loop, $c$, we see from Fig.~\ref{meanc} the interesting result that the average value $\langle c\rangle$ shows a linear behaviour (at least for $N$ ranging from 1 to 21), satisfying $\langle c \rangle  \simeq 2N$. We note that this result differs from that predicted in \cite{Copi:2010jw}, where it was suggested that $c \propto N^2$. The argument for $N^2$ is based on the fact that each mode roughly corresponds to a great circle on the Kibble-Turok sphere, so the number of intersections (i.e. the number of occasions a cusp forms) is proportional to $N^2$. It is not obvious to us at present why we are differing on this point; it is a very interesting question requiring further investigation. 
Figure~\ref{histogramc} depicts the frequency distribution of the cusps per period produced from the class of loops represented in Fig.~\ref{histogram}. It allows us to observe 
a general pattern of how the histogram changes with harmonic order. From the symmetries of the odd-harmonic string we conclude that the number of cusps per period, $c$, has to be an even number. Also, note that in low harmonics it is far more likely to have values of $c$ that are not multiplies of 4, as is clear from Fig.~\ref{histogramc}(a). This can be compared with the Turok solution \cite{Turok:1984cn}, where $c$ takes the value 4 only for very specific choices of the string parameters, and otherwise it takes the values 2 or 6 (see \cite{Anderson:2003gg}). As the harmonic order increases, we notice that it becomes more likely to have values of $c$ that are multiples of 4. It is worth noting a few interesting points from Figures \ref{meanc} and \ref{histogramc}. First of all note that even for low harmonic loops with $N=3$ there are on average 6 cusps per period, going up to approximately 40 for the case $N=19$. For the 1000 $N=3$ loops shown in Fig.~\ref{histogramc}, 650 of them have more than 6 cusps per period, for the $N=5$ case, that number rises to 850, and for the $N=19$ case, almost all the loops satisfy that condition. This raises the obvious question: what is the influence of these very cuspy loops when it comes to estimating the gravitational wave beaming from them? At the very least, it suggests that the effective number of cusps per period could well be significantly more than what has been assumed to date.

Table \ref{table}, provides an elegant summary of our main results. In it we show the values of the mean number of cusps per period for different harmonic order, as seen in Fig.~(\ref{meanc}). We also show the average values of $g_1$ and $g_2$ (including the maximum value we obtain for $g_1$) for each harmonic order. The take home message is pretty clear; the `cusp sharpness' parameters $g_1$ and $g_2$  are fairly closely distributed around unity, as has been assumed in the literature, and we have now demonstrated this assumption is justified. In fact, if anything they are slightly lower than unity, indicating the gravitational signals will be somewhat less strong from these effects (as far as $g_1$ is concerned), but mildly enhanced as far as $g_2$ is concerned, than has been assumed. On the other hand, as we discussed above, the average number of cusps per period, $c$, could be significantly enhanced for high-harmonic strings compared to the usual assumption $c\sim 1$, potentially leading to an enhancement of the predicted gravitational wave signal from string cusps.   

\begin{figure}%
    \centering
    \subfloat[]{{\includegraphics[scale=0.5]{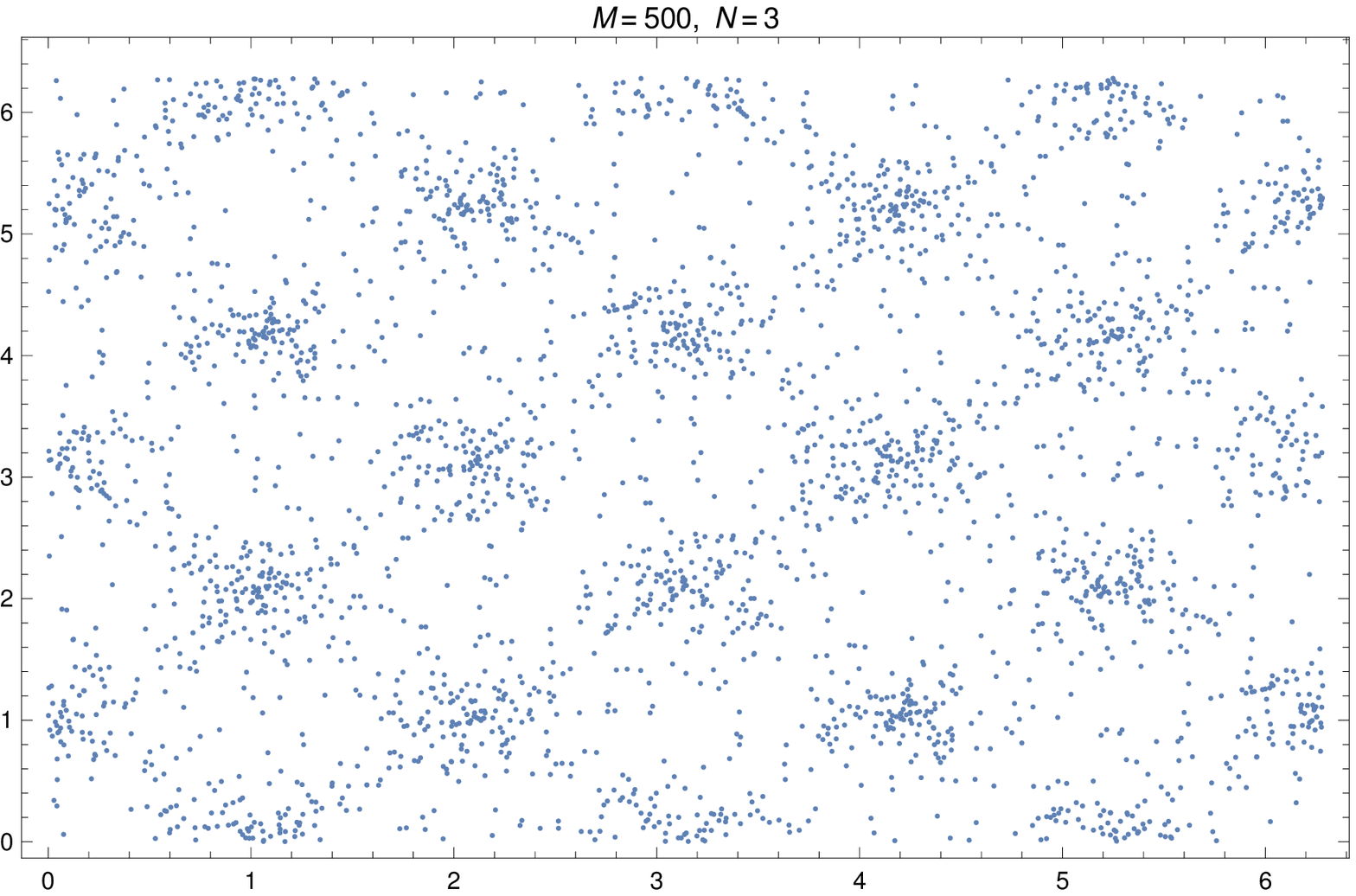} }}%
    \qquad
    \subfloat[]{{\includegraphics[scale=0.5]{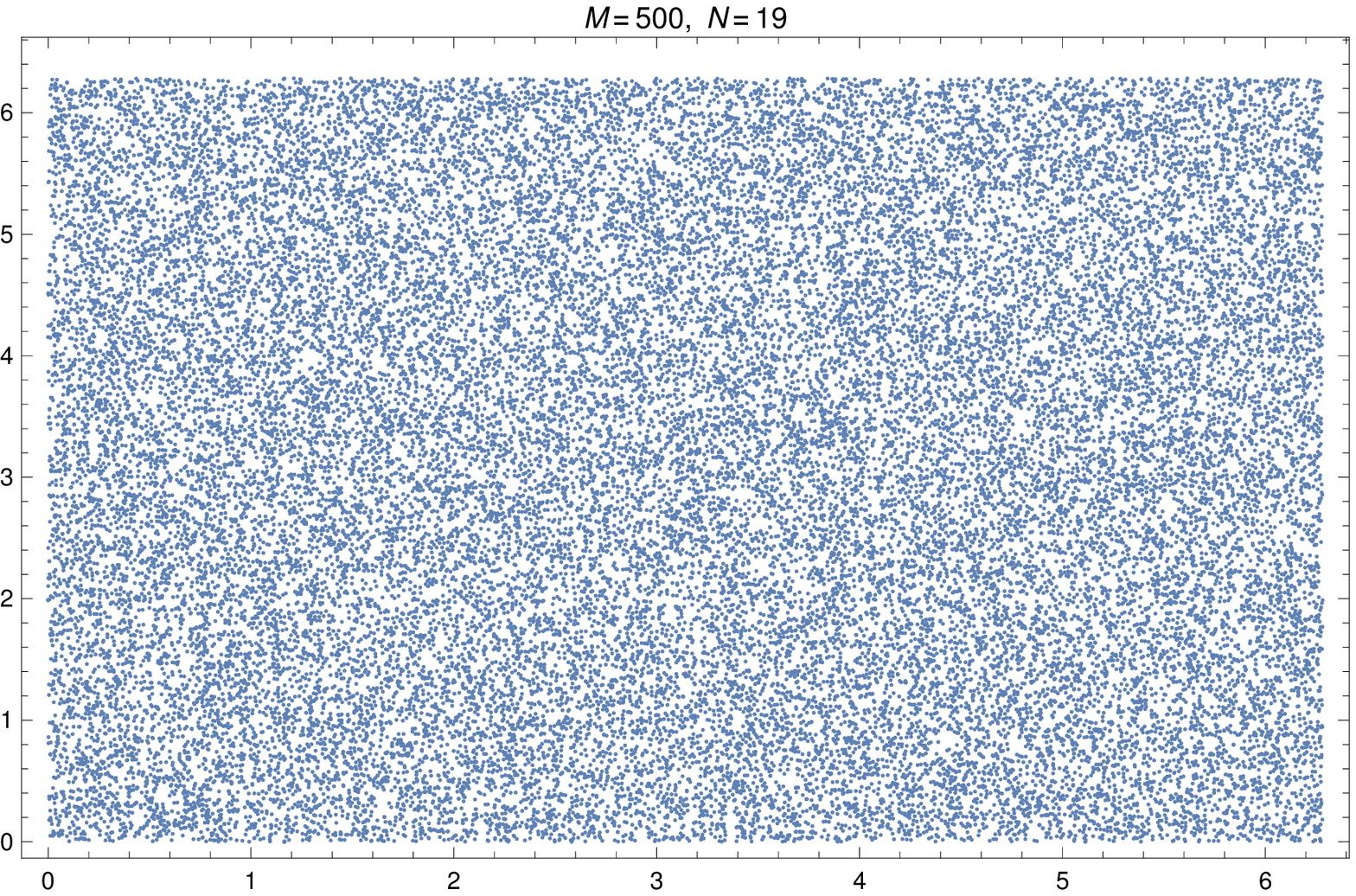} }}%
    \caption{The values of the $(u_c,v_c)$ pairs on the u-v plane. Note that for N=3 the pairs are more orderly placed, which is not observed for N=19.}%
    \label{uvpairs}%
\end{figure}

There is a nice formal mathematical aspect to the distribution of the $(u_c,v_c)$ pairs on the plane. In Fig.~\ref{uvpairs} we can see that the pairs of harmonic order N=3 follow a pattern, which does not persist for the higher harmonic orders, as we can see for the N=19 case. We can quantify this using the two-dimensional Kolmogorov-Smirnov test, which shows that for large harmonic orders (we have tested this up to N=19), the hypothesis that the distribution of the $(u_c,v_c)$ pairs follows the two-dimensional uniform distribution is not rejected at the 5 percent level (and so the distribution is consistent with being uniform). However, for harmonic order N=13 and smaller, we find that the uniform distribution hypothesis is rejected at the 5 percent level, indicating there is some underlying structure present. Another way to test the behaviour of the $(u_c,v_c)$ pairs is to convert the two-dimensional distribution to a one-dimensional one. One way to achieve this is to split the domain $[0,2\pi)\times [0,2\pi)$ in $u-v$ space into equal sized squares. The number of the squares is taken to be of the order of the number of $(u_c,v_c)$ pairs. We can then make a distribution of the number of $(u_c,v_c)$ pair counts in each square, and compare it with the poisson distribution of the same mean value. Using the Kolmogorov-Smirnov test to compare these two distributions we find again that the null hypothesis is rejected for N=13 or smaller at the 5 percent level, while it is not-rejected for N=15 to N=19, which is the maximum harmonic order we have tested. 

\begin{table}
\centering
\begin{tabular}{ |c||c|c|c|c|  }
 \hline
 Harmonic & c & $\langle g_1\rangle$ &$\text{Max}\left(\{ g_1\} \right)$ & $\langle g_2 \rangle$\\
 order & & & &\\
 \hline
 1 &  2.00 &  1.00 &  1.00 & 1.00 \\
 3 &  5.96 & 0.668 & 5.18 & 0.666 \\
 5 &  10.1 & 0.577 & 3.15 & 0.615 \\
 7 &  14.3 & 0.513 & 3.17 & 0.513 \\
 9 &  18.9 & 0.474 & 3.27 & 0.456 \\
11 &  22.6 & 0.446 & 3.01 & 0.432 \\
13 &  27.4 & 0.428 & 2.04 & 0.402 \\
15 &  30.4 & 0.400 & 3.35 & 0.374 \\
17 &  35.2 & 0.392 & 2.14 & 0.361 \\
19 &  39.0 & 0.375 & 1.91 & 0.341 \\

 \hline
\end{tabular}
\caption{Mean values of the cusp number per period and the average values of $g_1$ and $g_2$ evaluated at the cusp events for each harmonic order. We also provide the maximum value in the list of the values of $g_1$ we have obtained, $\{ g_1\}$, for each harmonic order.}
\label{table}
\end{table}

\section{Conclusions}\label{concl}
Cosmic strings have long been a favourite member of the particle cosmology family. Their formation out of symmetry breaking transitions in the early universe placed them as excellent candidates to play a role in the physics of structure formation. However, as the data began to role in, the evidence for such strings forming in a GUT type transition failed to turn up; strings were not putting in an appearance in the cosmic microwave background anisotropies. It meant that interest in them wained, but that is not quite the same as the strings not being present or relevant. The breakthrough discovery of gravitational waves from the merger of a pair of black holes \cite{Abbott:2016blz}, has transformed the field and given hope to the possibility that cosmic strings may yet be detected by their emission of gravitational waves, either through the stochastic gravitational wave background or through bursts of gravitational waves emitted from cusps and kinks on a network  \cite{Damour:2001bk, Damour:2004kw, Abbott:2017mem}. Concentrating on three models for the loop production function, the lack of evidence of any signal of gravitational wave bursts from cusps and kinks associated with string loops allowed the authors of \cite{Abbott:2017mem} to set upper limits on some of the key cosmic string parameters such as bounds on $G\mu$.  However, in reaching those conclusions they had to assume values for a number of key parameters, such as the average number of cusps formed per period on stable loop configurations, $c$, (called $N_q$ in \cite{Abbott:2017mem}) and the average value of the product of the second derivatives of the left and right moving vectors on the loop evaluated at the cusp, to be precise $g_1$ and $g_2$ defined in Eqs.~(\ref{g1-def}) and (\ref{g2-def}). These parameters which depend on the individual loops were both taken to be of order unity in \cite{Abbott:2017mem} and associated papers, and yet the final result for the amplitude of the gravitational wave signal from the bursts depends on them. If $c$ doubles then so does the amplitude, if $g_1$ increases by a factor of 2, then once again the signal doubles. 

It was this uncertainty in these key parameters that has motivated us to consider in this work the dynamics of cosmic string loops with higher harmonics. By going to higher harmonics we can first of all compare the cusp distribution to those of lower harmonic loops, but crucially we can gain excellent statistics on the range of values of $\vec{a}_N''$ and $\vec{b}_N''$, hence on the range of values of $g_1$ and $g_2$ associated with loops. Moreover, we have no way of a priori of estimating the distribution of loops formed in the early universe, nor the distribution of the initial large parent loops chopped off a long string. In fact it is quite likely they will be formed with many harmonics and begin oscillating with all of them in action. Given that, we have analysed a class of such loops, albeit with odd harmonics, first demonstrated in the elegant work of Siemens and Kibble \cite{Siemens:1994ir}. For simplicity we concentrated on the case where the left and right moving waves on the loop had the same number of harmonics, and for each harmonic we analysed thousands of loops generated at random. This was done for all the odd harmonics up to $N=21$. Two key results emerged. The first was that the average value for $g_1$ (and $g_2$) remained close to unity for all the harmonic cases studied, ranging from 1 (low harmonics) to 0.4 (high harmonics). In particular there were very few cases where it went above unity, indicating that the assumptions made about its behaviour (i.e. $g_1 \sim 1,\,g_2 \sim 1$) in \cite{Damour:2001bk} and \cite{Abbott:2017mem} is correct. The second concerned the parameter $c$ ($N_q$ in \cite{Abbott:2017mem}), the average number of cusps formed per period. Not surprisingly we found that it grew with harmonic number $N$, but rather than growing as $N^2$ as argued for in \cite{Copi:2010jw}, we found that it grew linearly obeying roughly $c = 2 N$ as seen in Fig.~\ref{meanc} and Table \ref{table}. Given that the amplitude of the gravitational wave signal from bursts from cusps is proportional to $c$, then this could be a significant feature. There is of course a caveat and that is the bounds of \cite{Damour:2001bk} and \cite{Abbott:2017mem} are based on the assumption that the loops being considered are non-self intersecting. The loops we are considering here can be considered as the initial loops formed from the long string network that will no doubt self intersect within the first oscillation. Such loops are constantly chopping off the network, and if there are a large number of cusps on these initial large loops they may well modify the burst output in those opening oscillations compared to the loops which are assumed to have just one cusp per cycle. We will be addressing the impact of these loops on the gravitational wave signatures constrained by LIGO in a future publication \cite{inprog}, as well as modelling the process of self intersection of these high harmonic initial loops. 

\acknowledgements
We would like to thank Dani Steer, Christophe Ringeval, Ken Olum, Jose Blanco-Pillado and Konstantinos Palapanidis for very useful conversations. Two of us (AA and EJC) are very grateful to  Ana Achucarro, Leandros Perivolaropoulos, Tanmay Vachaspati, and the Lorentz Center for organising the workshop ``Cosmic Topological Defects: Dynamics and Multi-messenger Signatures", where some of this work was discussed. AA and EJC acknowledge support from STFC grant ST/P000703/1. DP acknowledges support from the University of Nottingham Vice Chancellor's Scholarship.

\end{document}